\documentclass[11pt,preprintnumbers,titlepage]{revtex4-2}%
\usepackage[latin9]{inputenc}
\usepackage{amsmath}
\usepackage{amssymb}
\usepackage{graphicx}
\usepackage[unicode=true,  bookmarks=false,  breaklinks=false,pdfborder={0 0
1},backref=section,colorlinks=false]{hyperref}
\usepackage{amsfonts}
\usepackage{subfigure}
\usepackage{cleveref}
\usepackage{listings}
\usepackage{xcolor}
\usepackage{array}
\usepackage{url}
\usepackage{color}%
\hypersetup{
colorlinks,linkcolor=blue,citecolor=blue,urlcolor=blue}
\makeatletter
\@ifundefined{textcolor}{}{
\definecolor{BLACK}{gray}{0}
\definecolor{WHITE}{gray}{1}
\definecolor{RED}{rgb}{1,0,0}
\definecolor{GREEN}{rgb}{0,1,0}
\definecolor{BLUE}{rgb}{0,0,1}
\definecolor{CYAN}{cmyk}{1,0,0,0}
\definecolor{MAGENTA}{cmyk}{0,1,0,0}
\definecolor{YELLOW}{cmyk}{0,0,1,0}
}

\makeatother
\begin{document}
\preprint{CTP-SCU/2022008}
\title{Probing Phase Structure of Black Holes with Lyapunov Exponents}
\author{Xiaobo Guo$^{b}$}
\email{guoxiaobo@czu.edu.cn}
\author{Yuhang Lu$^{c}$}
\email{luyuhang668@stu.scu.edu.cn}
\author{Benrong Mu$^{a}$}
\email{benrongmu@cdutcm.edu.cn}
\author{Peng Wang$^{c}$}
\email{pengw@scu.edu.cn}
\affiliation{$^{a}$Physics Teaching and Research Section, College of Medical Technology,
  Chengdu University of Traditional Chinese Medicine, Chengdu, 611137, PR China}
\affiliation{$^{b}$Mechanical and Electrical Engineering School, Chizhou University,
  Chizhou, Anhui, 247000, PR China }
\affiliation{$^{c}$Center for Theoretical Physics, College of Physics, Sichuan University,
  Chengdu, 610064, PR China}

\begin{abstract}
  We conjecture that there exists a relationship between Lyapunov exponents and
  black hole phase transitions. To support our conjecture, Lyapunov exponents of
  the motion of particles and ring strings are calculated for
  Reissner-Nordstr\"{o}m-AdS black holes. When a phase transition occurs, the
  Lyapunov exponents become multivalued, and branches of the Lyapunov exponents
  coincide with black hole phases. Moreover, the discontinuous change in the
  Lyapunov exponents can be treated as an order parameter, and has a critical
  exponent of $1/2$ near the critical point. Our findings reveal that Lyapunov
  exponents can be an efficient tool to study phase structure of black holes.

\end{abstract}
\maketitle
\tableofcontents

{}

{}

\section{Introduction}

Black hole thermodynamics lies in the interdisciplinary area of general
relativity, quantum mechanics, information theory and statistical physics, and
can provide profound insights into the nature of gravity. The area theorem,
which asserts that the total horizon area of black holes is a non-decreasing
function of time \cite{Hawking:1971tu}, suggests that black holes may be
endowed with thermodynamic properties. Inspired by the resemblance between the
area theorem and the second law of thermodynamics, Bekenstein postulated that
black hole entropy can be described by the horizon area
\cite{Bekenstein:1972tm,Bekenstein:1973ur}. The analogy between usual
thermodynamics and black hole thermodynamics was further enhanced by the
discovery of Hawking radiation, assigning black holes a temperature
\cite{Hawking:1974rv,Hawking:1975iha}.

Later on, Hawking and Page discovered that there exists a phase transition
between Schwarzschild-AdS black holes and a thermal space
\cite{Hawking:1982dh}. With the advent of the AdS/CFT correspondence
\cite{Maldacena:1997re,Gubser:1998bc}, thermodynamics and phase structure of
various AdS black holes have been widely studied
\cite{Witten:1998qj,Witten:1998zw,Cvetic:1999ne,Chamblin:1999tk,Chamblin:1999hg,Caldarelli:1999xj,Cai:2001dz,Cvetic:2001bk,Nojiri:2001aj}%
. Specifically, Reissner-Nordstr\"{o}m-AdS (RN-AdS) black holes exhibit a van
der Waals-like phase transition, which consists of a first-order phase
transition terminating at a second-order critical point, in a canonical
ensemble \cite{Chamblin:1999tk,Chamblin:1999hg}, and a Hawking-Page-like phase
transition in a grand canonical ensemble \cite{Cai:2001dz}. In the extended
phase space with the cosmological constant being treated as a thermodynamic
pressure \cite{Kastor:2009wy,Dolan:2011xt,Kubiznak:2012wp}, phase behavior and $P$-$V$
criticality have been explored for AdS black holes, which discovered a broad
range of new phenomena
\cite{Wei:2012ui,Gunasekaran:2012dq,Cai:2013qga,Altamirano:2013ane,Altamirano:2013uqa,Xu:2014kwa,Frassino:2014pha,Dehghani:2014caa,Wei:2014hba,Dolan:2014vba,Hennigar:2015esa,Caceres:2015vsa,Wei:2015ana,Chakraborty:2015hna,
  Hendi:2016yof,Hennigar:2016xwd,Momeni:2016qfv,Hendi:2017fxp,Lemos:2018cfd,Pedraza:2018eey,Wang:2018xdz,Wei:2020poh}%
. In the extended phase space, the analogy between RN-AdS black holes and the
van der Waals fluid becomes more complete, in that the coexistence lines in
the $P$-$T$ diagram are both finite and terminate at critical points, and the
$P$-$V$ criticality matches with one another \cite{Kubiznak:2012wp}.

Since the nature of black hole thermodynamics has not yet been fully
understood, it is of great interest to explore phase structure of black holes
from various perspectives. For example, the Ruppeiner geometry can be
exploited to probe the microstructure of black holes
\cite{Ruppeiner:2012uc,Miao:2017cyt,Guo:2019oad,Wei:2019yvs,Wang:2019cax,Yerra:2020oph,Yerra:2021hnh}.
Motivated by the Ruppeiner geometry, RN-AdS black holes have been proposed to
be built of some unknown micromolecules \cite{Wei:2015iwa,Wei:2019uqg}. More
interestingly, there have been attempts to associate phase transitions of
black holes with some observational signatures, such as quasinormal modes
\cite{Liu:2014gvf,Mahapatra:2016dae,Chabab:2016cem,Zou:2017juz,Zhang:2020khz},
circular orbit radius of a test particle
\cite{Wei:2017mwc,Wei:2018aqm,Zhang:2019tzi} and black hole shadow radius
\cite{Zhang:2019glo,Belhaj:2020nqy}. It showed that phase structure of black
holes can be revealed by behavior of the aforementioned physical quantities,
and the discontinuity in the physical quantities across phase transitions
behaves similarly to an order parameter.

Lyapunov exponents characterize the rate of separation of adjacent
trajectories, and positive/negative Lyapunov exponents correspond to
divergent/convergent trajectories \cite{lyapunov1992general}. Lyapunov
exponents can be used to study chaotic dynamics in general relativity, which
is a nonlinear dynamical theory. The chaotic motion of particles in black hole
spacetime has been extensively studied, such as static axisymmetric spacetimes
\cite{Sota:1995ms,Sota:1996cv}, rotating charged black hole spacetimes \cite{Kan:2021blg,Gwak:2022xje}, multi-black hole spacetimes
\cite{Hanan:2006uf}, bumpy spacetimes \cite{Gair:2007kr}, weakly magnetized
Schwarzschild black holes \cite{Zahrani:2013up}, black holes with discs or
rings \cite{Polcar:2019kwu}, Schwarzschild-Melvin black holes
\cite{Wang:2016wcj}, accelerating black holes \cite{Chen:2016tmr}, spacetimes
with a quadrupole mass moment \cite{Wang:2018eui} and black holes with quantum
gravity corrections \cite{Lu:2018mpr,Guo:2020xnf}. Particularly, the motion of
a particle near the black hole horizon was studied in
\cite{Hashimoto:2016dfz,Dalui:2018qqv}, which found that the Lyapunov exponent
obeys an universal upper bound proposed in the framework of gauge/gravity
duality \cite{Maldacena:2015waa}. Nevertheless, counterexamples that violate
the upper bound have been reported \cite{Zhao:2018wkl,Guo:2020pgq}. Moreover,
partly motivated by gauge/gravity duality, chaotic dynamics of a ring string
has been studied in Schwarzschild-AdS and charged AdS black holes
\cite{PandoZayas:2010xpn, Ma:2014aha, Basu:2016zkr, Hashimoto:2018fkb,
  Cubrovic:2019qee,Ma:2019ewq,Ma:2022tvs}. Intriguingly, Lyapunov exponents of
unstable null geodesics have been revealed to be closely related to the
imaginary part of a class of quasinormal modes of perturbations in black hole
spacetime \cite{Cardoso:2008bp,Guo:2021enm}.

In this paper, we aim to explore the relationship between phase structure of
RN-AdS black holes and Lyapunov exponents of particles and ring strings moving
in the black holes. The rest of this paper is organized as follows.
Thermodynamics and phase transitions of RN-AdS black holes are briefly
reviewed in Sec. \ref{sec:pt}. Focusing on particles, we examine the
relationship between Lyapunov exponents of unstable circular geodesics and
phase structure of RN-AdS black holes in Sec. \ref{sec:particle}. The case of
a ring string is discussed in Sec. \ref{sec:string}. We summarize our results
with a brief discussion in Sec. \ref{sec:conclusion}. For simplicity, we set
$G=\hbar=k_{B}=c=1$ in this paper.

\section{Phase Structure of RN-AdS Black Holes}

\label{sec:pt}

In this section, we review thermodynamic properties and phase structure of
RN-AdS black holes. The 4-dimensional static charged RN-AdS black hole
solution is described by
\begin{equation}
  ds^{2}=-f\left(  r\right)  dt^{2}+\frac{1}{f\left(  r\right)  }dr^{2}%
  +r^{2}\left(  d\theta^{2}+\sin^{2}\theta d\varphi^{2}\right)  ,
\end{equation}
where the metric function $f(r)$ is
\begin{equation}
  f\left(  r\right)  =1-\frac{2M}{r}+\frac{Q^{2}}{r^{2}}+\frac{r^{2}}{l^{2}},
\end{equation}
and $l$ is the AdS radius. Here, the parameters $M$ and $Q$ can be interpreted
as the black hole mass and charge, respectively. The RN-AdS black hole has an
event horizon at $r=r_{+}$, and the horizon radius $r_{+}$ satisfies
$f(r_{+})=0$. In terms of $r_{+}$, the Hawking temperature $T$ and the mass
$M$ are given by \cite{Wang:2019cax}
\begin{equation}
  T=\frac{1}{4\pi r_{+}}\left(  1-\frac{Q^{2}}{r_{+}^{2}}+\frac{3r_{+}^{2}%
  }{l^{2}}\right)  ,M=\frac{r_{+}}{2}\left(  1+\frac{Q^{2}}{r_{+}^{2}}%
  +\frac{r_{+}^{2}}{l^{2}}\right)  , \label{eq:tem}%
\end{equation}
respectively. Moreover, the RN-AdS black hole obeys the first law of
thermodynamics,
\begin{equation}
  dM=TdS+\Phi dQ,
\end{equation}
where $S=\pi r_{+}^{2}$ and $\Phi=Q/r_{+}$ are the entropy and the potential
of the black hole, respectively. By computing the Euclidean action in the
semiclassical approximation, we obtain the free energy
\begin{equation}
  F=M-TS=\frac{1}{4}\left(  \frac{3Q^{2}}{r_{+}}+r_{+}-\frac{r_{+}^{3}}{l^{2}%
  }\right)  . \label{F}%
\end{equation}
By dimensional analysis, we find that the physical quantities scale as powers
of $l$,
\begin{equation}
  \tilde{Q}=Q/l,\ \tilde{r}_{+}=r_{+}/l,\ \tilde{T}=Tl,\ \tilde{F}%
  =F/l,\ \tilde{M}=M/l,\ \tilde{r}=r/l, \label{rescale}%
\end{equation}
where the tildes denote dimensionless quantities. \begin{figure}[pt]
  \begin{center}
    \includegraphics[width=0.48\textwidth]{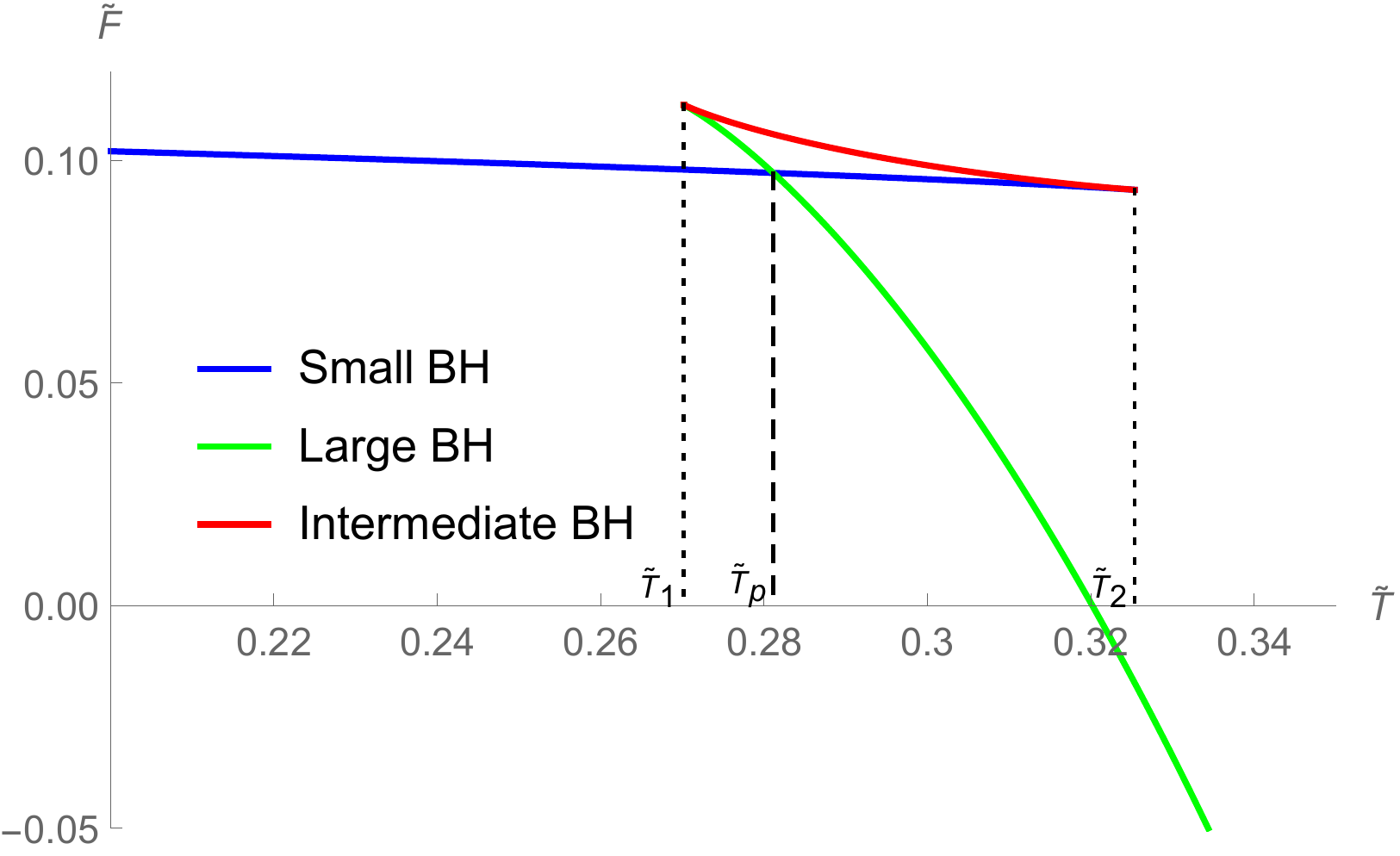} \hspace{1em}
    \includegraphics[width=0.48\textwidth]{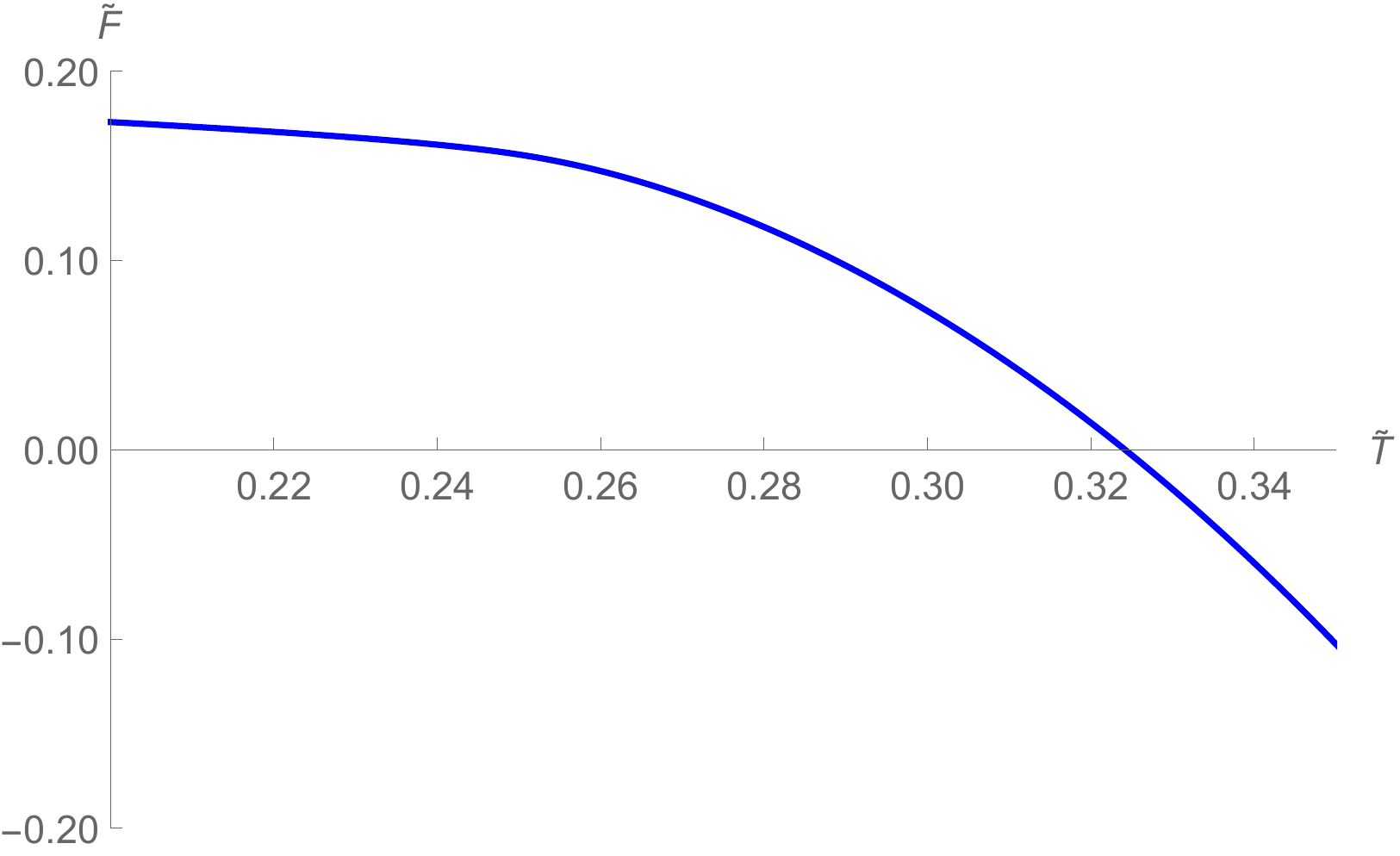}
  \end{center}
  \caption{Free energy $\tilde{F}$ as a function of the temperature $\tilde{T}$
    for RN-AdS black holes. \textbf{Left:} $\tilde{Q}=0.11<\tilde{Q}_{c}$. There
    coexist three black hole solutions when $\tilde{T}_{1}<\tilde{T}<\tilde{T}%
      _{2}$, and a first-order phase transition between Small BH and Large BH occurs
    at $\tilde{T}_{p}$. \textbf{Right:} $\tilde{Q}=0.20>\tilde{Q}_{c}$. Only one
    black hole solution exists, and hence no phase transition occurs.}%
  \label{pt}%
\end{figure}

Using eqn. $\left(  \ref{eq:tem}\right)  $, we can express $\tilde{r}_{+}$ as
a function of $\tilde{T}$. If $\tilde{r}_{+}(\tilde{T})$ is multivalued, there
is more than one black hole solution for fixed values of $\tilde{Q}$ and
$\tilde{T}$, corresponding to multiple phases in a canonical ensemble. The
critical point is an inflection point determined by
\begin{equation}
  \frac{\partial\tilde{T}}{\partial\tilde{r}_{+}}=0,\frac{\partial^{2}\tilde{T}%
  }{\partial\tilde{r}_{+}^{2}}=0,
\end{equation}
which gives the corresponding quantities evaluated at the critical point,
\begin{equation}
  \tilde{r}_{+c}=\frac{1}{\sqrt{6}},\ \tilde{Q}_{c}=\frac{1}{6},\ \tilde{T}%
  _{c}=\frac{1}{\pi}\sqrt{\frac{2}{3}},\ \Phi_{c}=\frac{1}{\sqrt{6}}.
\end{equation}
To study phase transitions, we express $\tilde{F}$ with respect to $\tilde{T}$
by plugging $\tilde{r}_{+}(\tilde{T})$ into eqn. $\left(  \ref{F}\right)  $
and plot $\tilde{F}$ against $\tilde{T}$ for various $\tilde{Q}$ in FIG.
\ref{pt}. The left panel shows that, when $\tilde{Q}<\tilde{Q}_{c}$, there are
three black solutions, dubbed as Small BH, Intermediate BH and Large BH. The
three black hole solutions coexist for some range of $\tilde{T}$, and a
first-order phase transition occurs at $\tilde{T}_{p}$. When $\tilde{Q}%
  >\tilde{Q}_{c}$, there is only one black hole solution and no phase
transition, which is shown in the right panel.

\section{Phase Transitions and Lyapunov Exponents of Particles}

\label{sec:particle}

In this section, we investigate the relationship between Lyapunov exponents of
massless and massive particles and phase transitions of RN-AdS black holes. In
particular, we focus on unstable circular geodesics on the equatorial
hyperplane with $\theta=$ $\pi/2$, which are described by the Lagrangian
\begin{equation}
  2\mathcal{L}=-f\left(  r\right)  \dot{t}+\frac{1}{f\left(  r\right)  }\dot
  {r}^{2}+\dot{r}^{2}\dot{\varphi}^{2}.
\end{equation}
Here dots and primes denote derivatives with respect to the proper time and
the areal radius $r$, respectively. Then the radial motion can be expressed
as
\begin{equation}
  \dot{r}^{2}+V_{\text{eff}}\left(  r\right)  =E^{2},
\end{equation}
where the constant $E$ can be treated as the energy and the energy per unit
mass for massless and massive particles, respectively. Here, we introduce the
effective potential,
\begin{equation}
  V_{\text{eff}}\left(  r\right)  =f(r)\left[  \frac{L^{2}}{r^{2}}+\delta
    _{1}\right]  ,
\end{equation}
where $L$ is identified as the angular momentum of the particles, and
$\delta_{1}=1$ and $0$ correspond to massless and massive particles,
respectively. The radius of an unstable circular geodesic is determined by
\begin{equation}
  \ V_{\text{eff}}^{\prime}\left(  r\right)  =0,\ V_{\text{eff}}^{\prime\prime
    }\left(  r\right)  >0.
\end{equation}

\subsection{Massless Particles}

\begin{figure}[ptb]
  \centering
  \includegraphics[width=8cm]{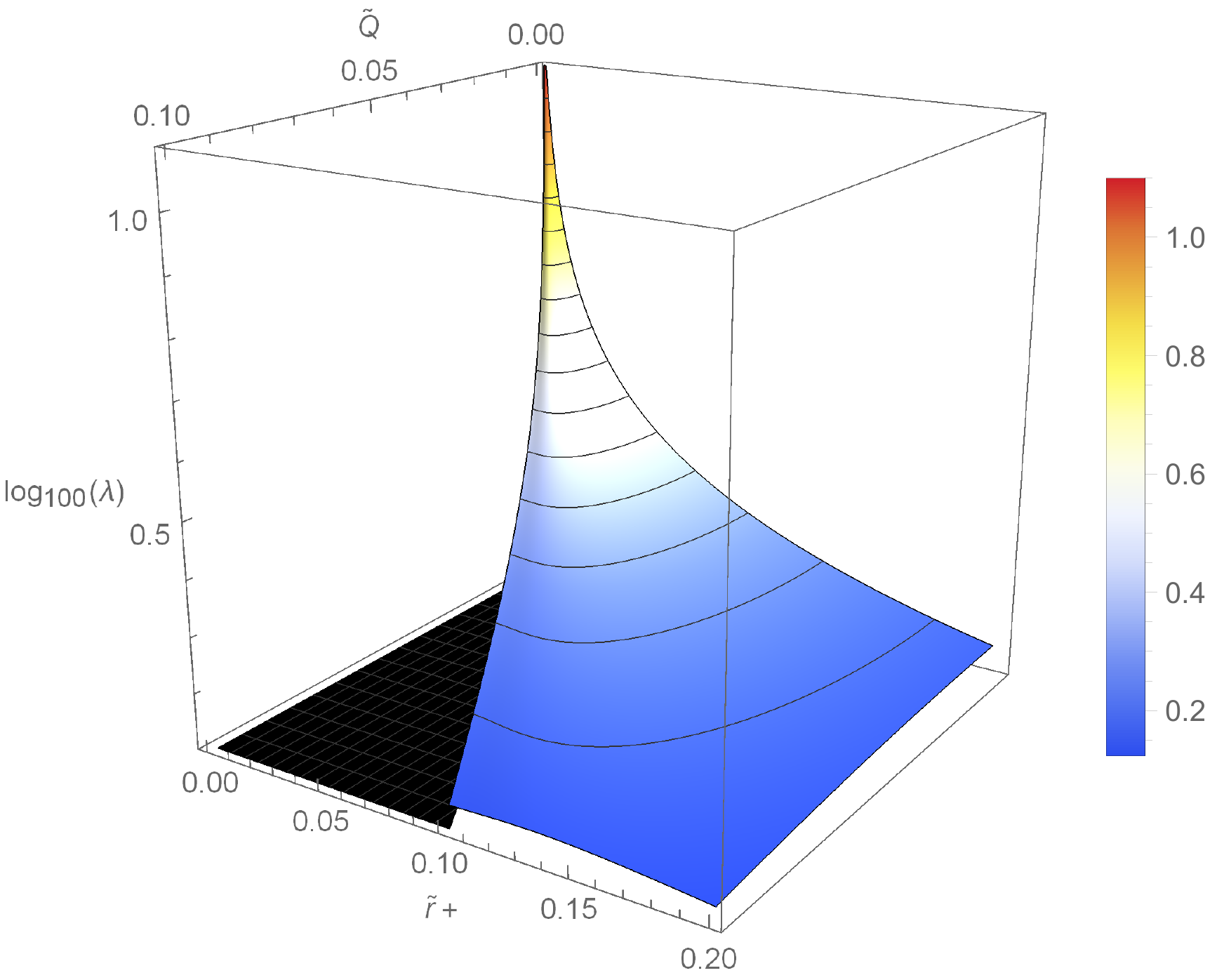} \caption{Three-dimensional plot of
    $\log_{100}\lambda$ as a function of $\tilde{Q}$ and $\tilde{r}_{+}$ for
    unstable null circular geodesics. Black hole solutions do not exist in the
    black region.}%
  \label{density1}%
\end{figure}

For massless particles, there is always (except for $L=0$) an unstable
circular geodesic outside the event horizon at%
\begin{equation}
  r_{\text{o}}=\frac{l}{2}\left[  \frac{3}{2}\tilde{r}_{+}\left(  \frac
    {\tilde{Q}^{2}}{\tilde{r}_{+}^{2}}+\tilde{r}_{+}^{2}+1\right)  +\sqrt{\frac
      {9}{4}\tilde{r}_{+}^{2}\left(  \frac{\tilde{Q}^{2}}{\tilde{r}_{+}^{2}}%
      +\tilde{r}_{+}^{2}+1\right)  ^{2}-8\tilde{Q}^{2}}\right]  , \label{rps}%
\end{equation}
which is independent of $L$. Furthermore, the Lyapunov exponent of the
unstable null circular geodesic is given by \cite{Cardoso:2008bp}
\begin{equation}
  \lambda=\sqrt{\frac{r_{\text{o}}^{2}f\left(  r_{\text{o}}\right)  }{L^{2}%
  }V_{\text{eff}}^{\prime\prime}\left(  r_{\text{o}}\right)  }, \label{lambda}%
\end{equation}
which depends only on $\tilde{Q}$ and $\tilde{r}_{+}$. We present the 3D plot
of $\log_{100}\lambda$ as a function of $\tilde{Q}$ and $\tilde{r}_{+}$ in
FIG. \ref{density1}, where black hole solutions with the event horizon at
$\tilde{r}_{+}$ do not exist in the black region. It shows that $\lambda$
diverges when $\tilde{r}_{+}$ decrease to zero. In addition, $\lambda$
approaches $1$ as $\tilde{Q}$ or $\tilde{r}_{+}$ increases to infinity. In
fact, eqns. $\left(  \ref{rps}\right)  $ and $\left(  \ref{lambda}\right)  $
indicate that, when $\tilde{r}_{+}$ or $\tilde{Q}$ approaches infinity,
$\tilde{r}_{\text{o}}$ goes to infinity, and hence $\lambda$ goes to $1$.

\begin{figure}[ptb]
  \begin{center}
    \includegraphics[width=0.40\textwidth]{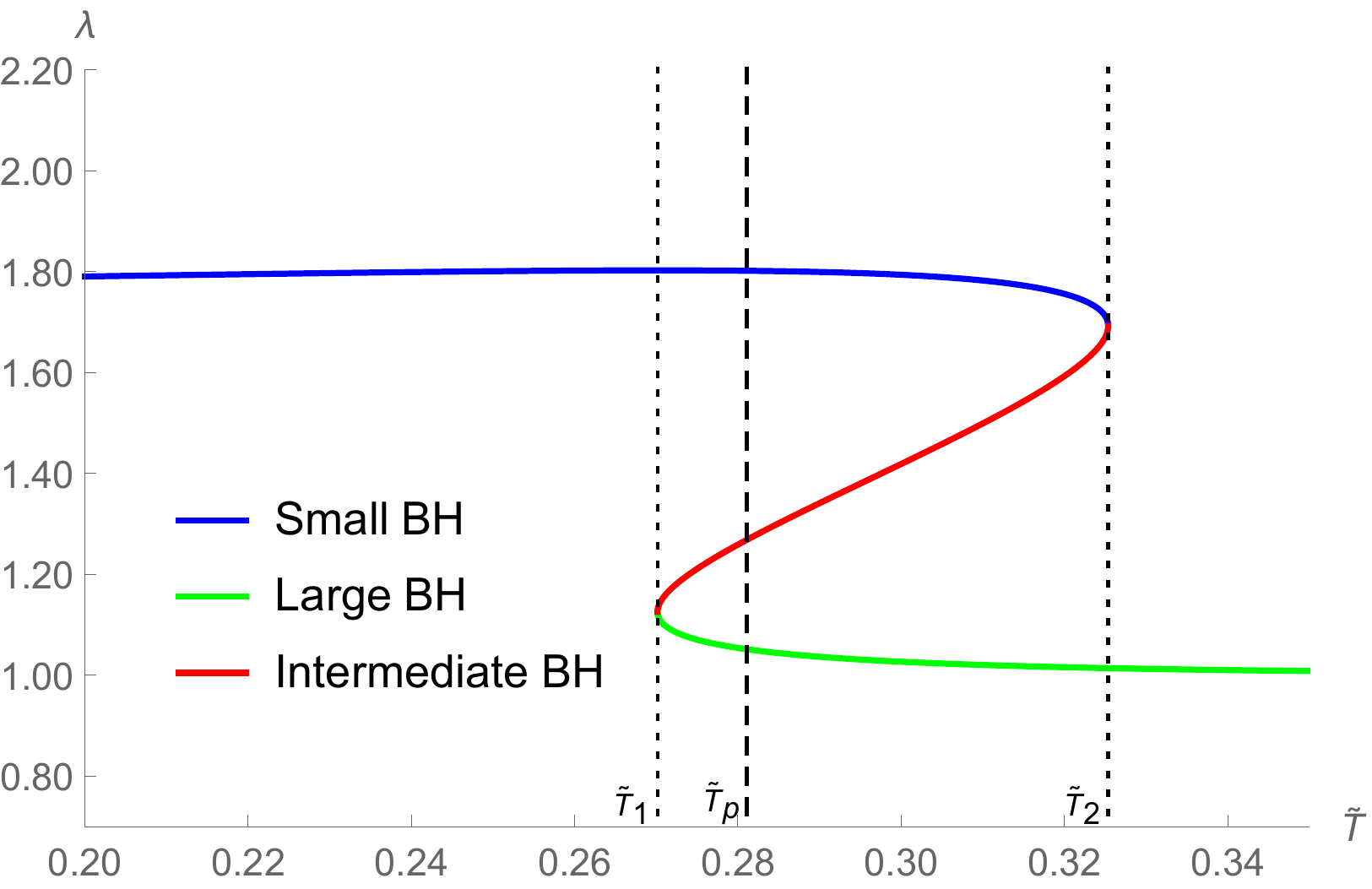} \hspace{1em}
    \includegraphics[width=0.40\textwidth]{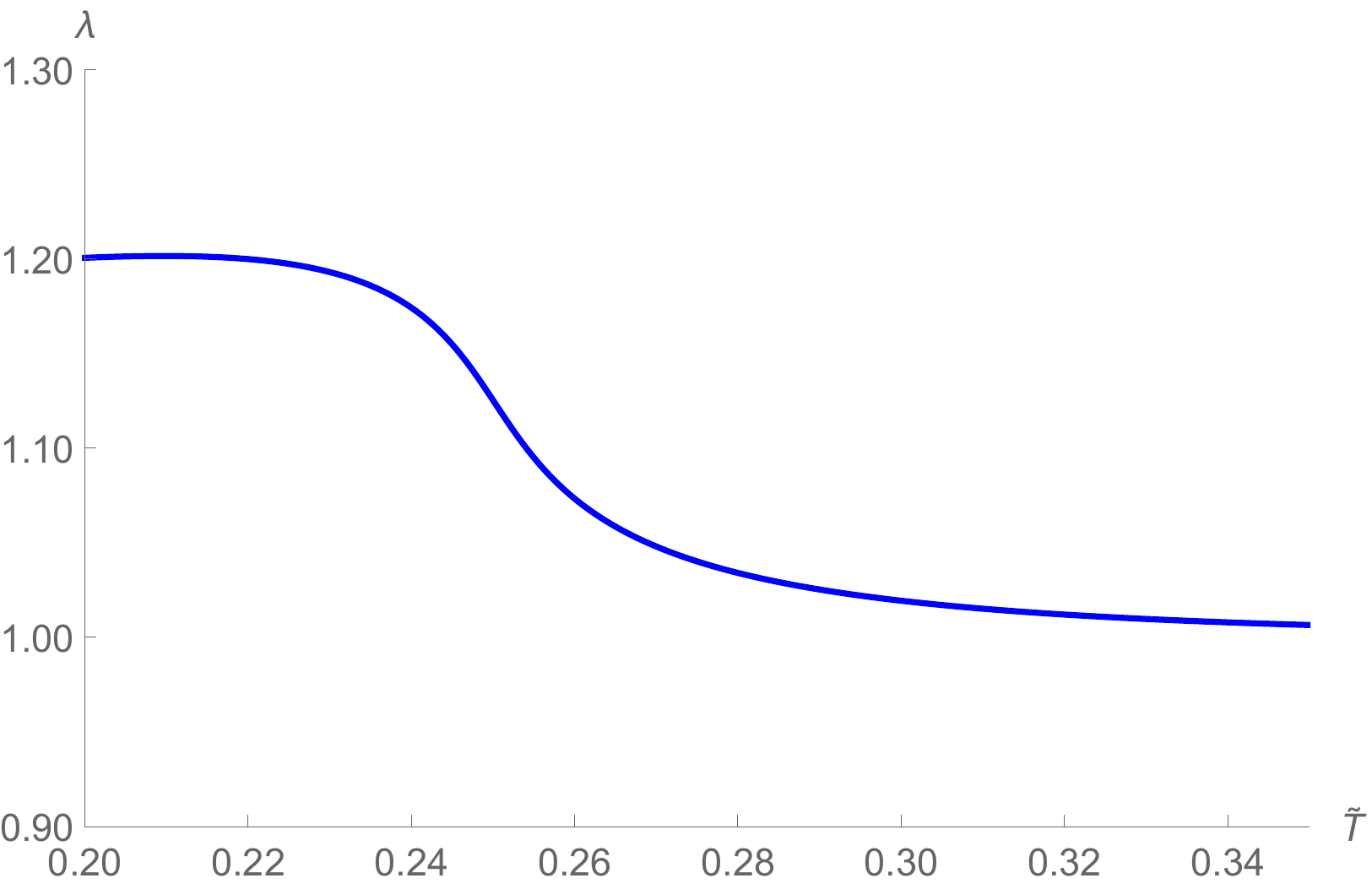}
    \includegraphics[width=0.40\textwidth]{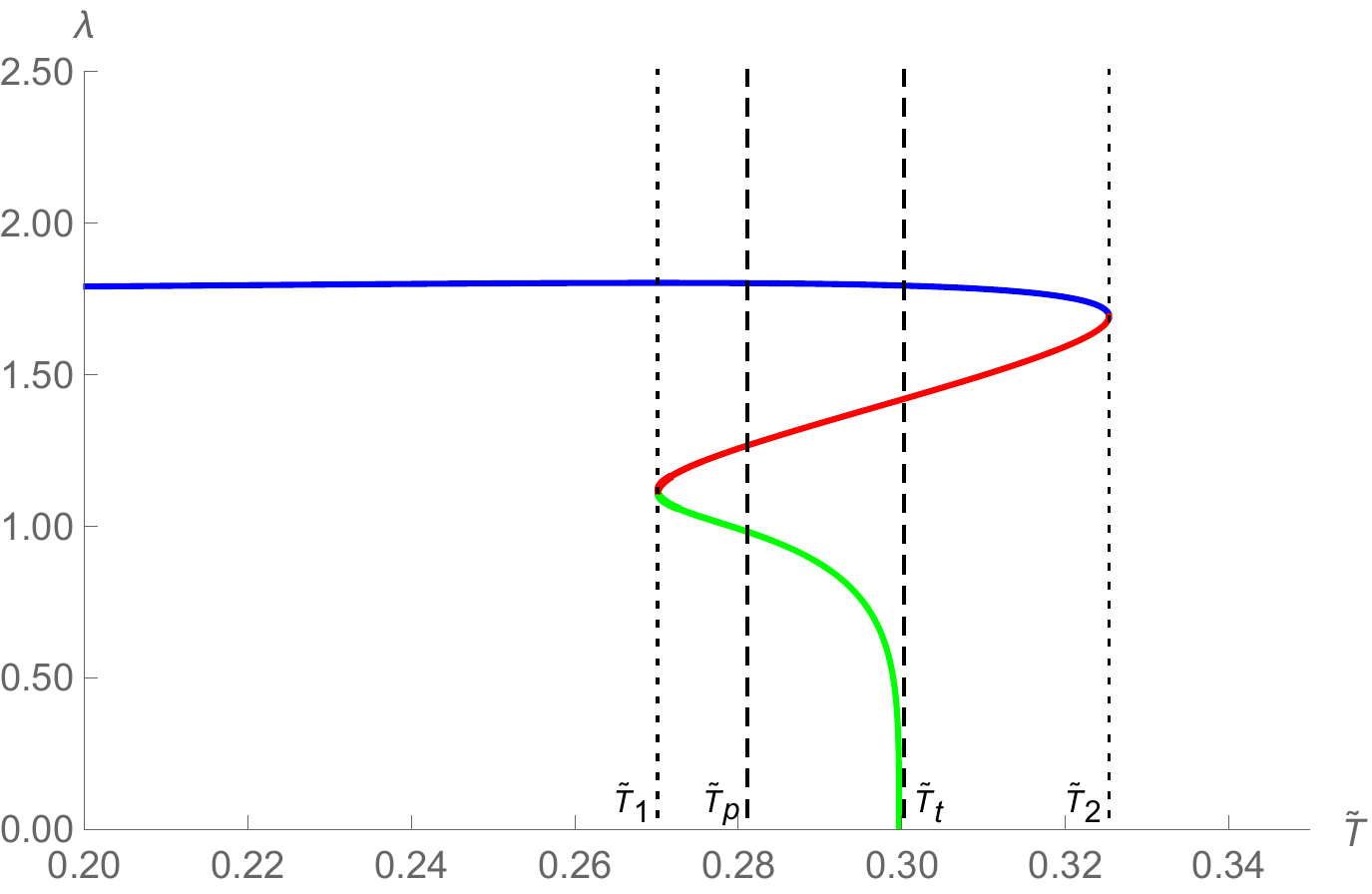} \hspace{1em}
    \includegraphics[width=0.40\textwidth]{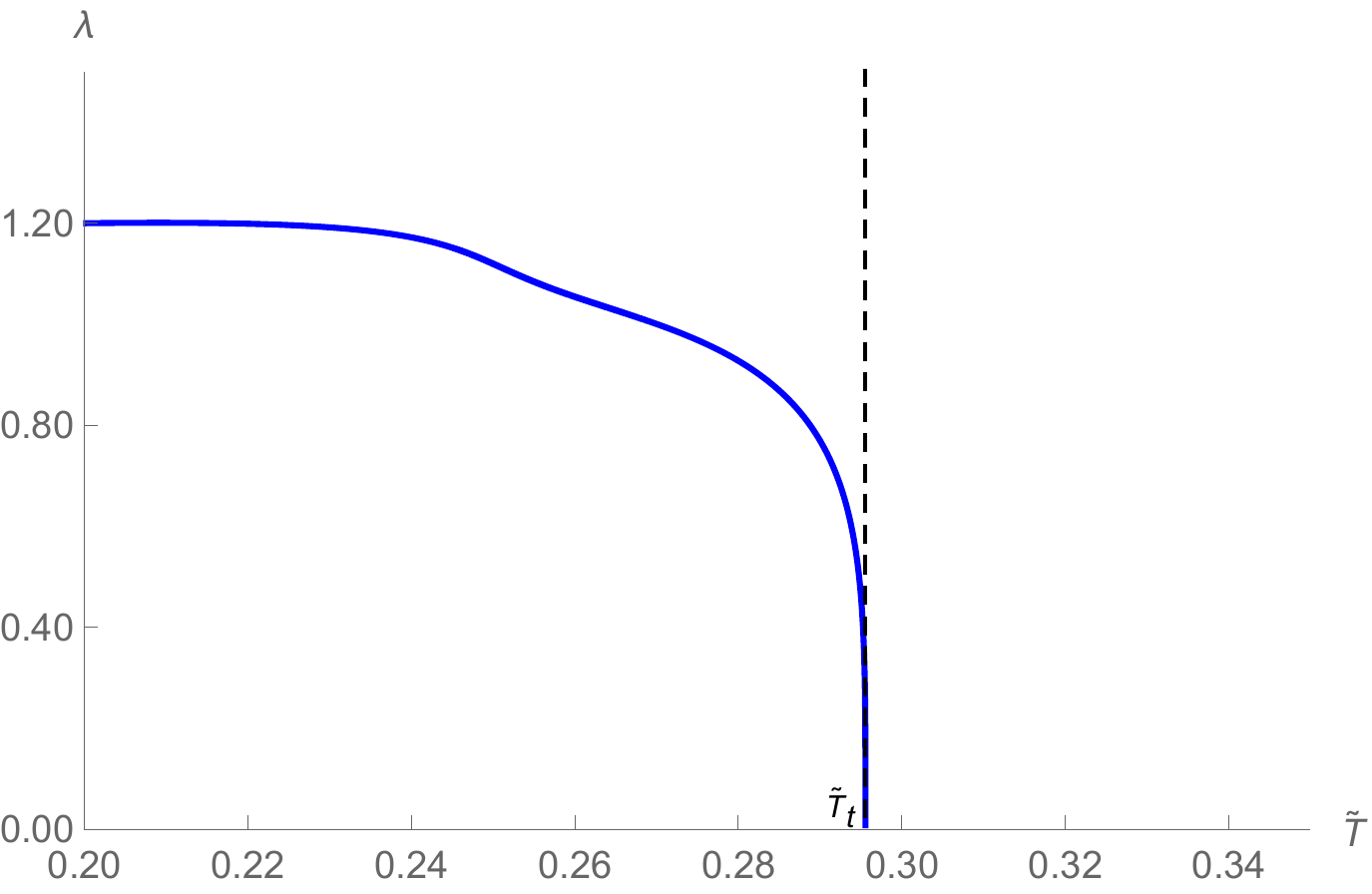}
    \includegraphics[width=0.40\textwidth]{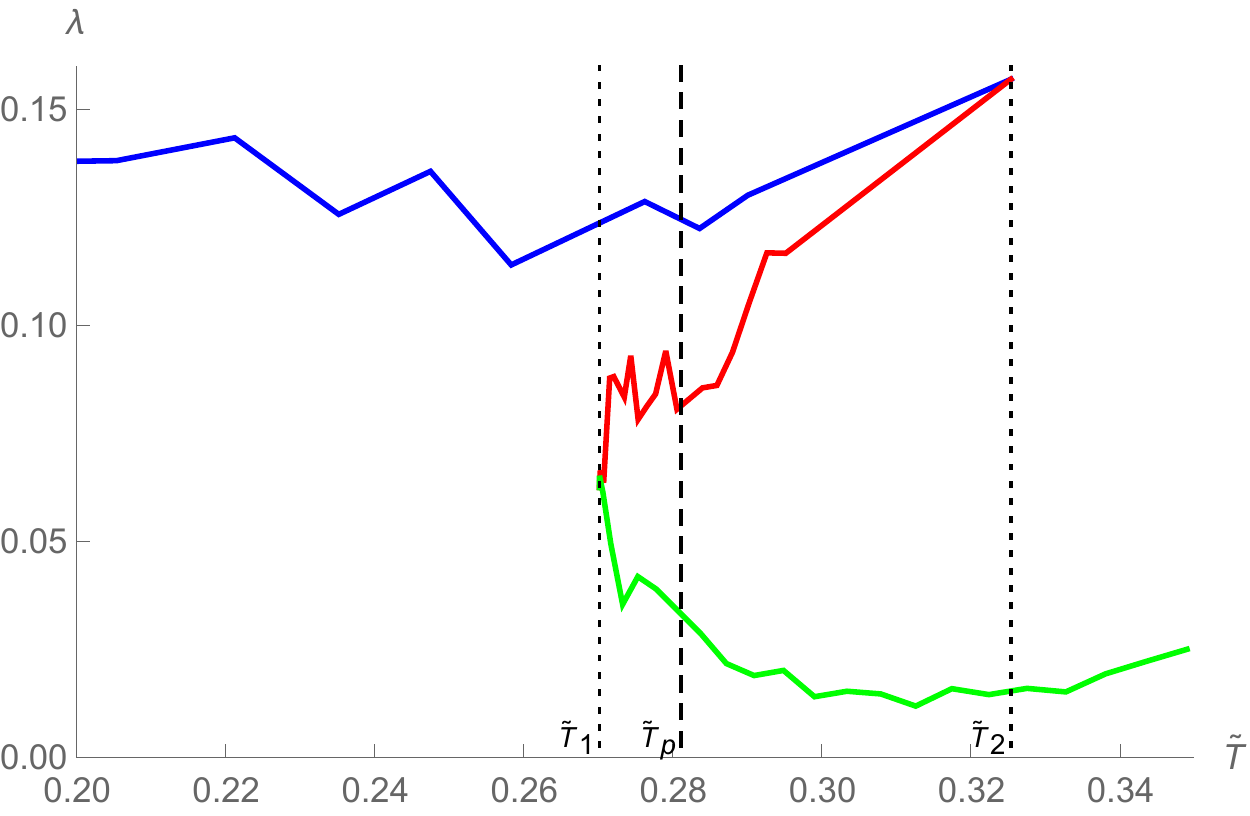} \hspace{1em}
    \includegraphics[width=0.40\textwidth]{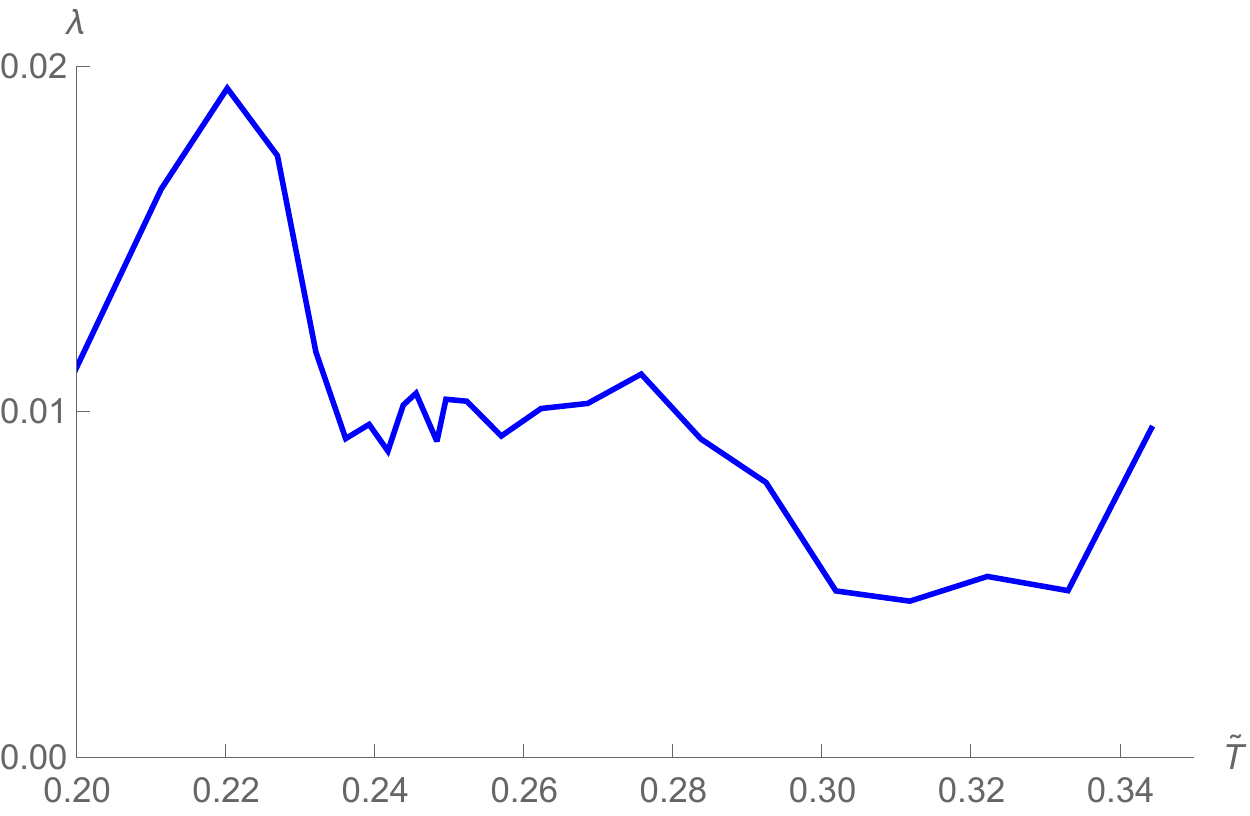}
  \end{center}
  \caption{Lyapunov exponents $\lambda$ of particles and strings as a function
    of the temperature $\tilde{T}$ for $\tilde{Q}=0.11<\tilde{Q}_{c}$
    (\textbf{Left Column}) and $\tilde{Q}=0.20>\tilde{Q}_{c}$ (\textbf{Right
      Column}). \textbf{Top Row}: Massless particles on the unstable null circular
    geodesic; \textbf{Middle Row}: Massive particles with $L=20l$ on the unstable
    time-like circular geodesic; \textbf{Bottom row}: Motion of ring strings with
    the initial conditions $\theta_{0}=0$, $E=1000$, $\frac{d}{d\tau}\left(
      r\cos\theta\right)  =0$ and $r_{0}=5.92$ (\textbf{Left}) and $7.30$
    (\textbf{Right}). Three black hole solutions, i.e., Small BH, Intermediate BH
    and Large BH, coexist for $\tilde{T}_{1}<\tilde{T}<\tilde{T}_{2}$. The phase
    transition between Small BH and Large BH occurs at $\tilde{T}=\tilde{T}_{p}$,
    and $\lambda$ of massive particles vanishes at $\tilde{T}=\tilde{T}_{t}$. When
    $\tilde{Q}<\tilde{Q}_{c}$, $\lambda$ is a multivalued function of $\tilde{T}$
    with three branches, which coincide with the three black hole solutions,
    respectively. When $\tilde{Q}>\tilde{Q}_{c}$, $\lambda$ is single-valued,
    demonstrating that there is only one black hole solution. These observations
    imply that $\lambda$ as a function of $\tilde{T}$ can reflect the phase
    structure of RN-AdS black holes.}%
  \label{fig:lambdaT}%
\end{figure}

Plugging $\tilde{r}_{+}(\tilde{T})$ into eqn. $\left(  \ref{lambda}\right)  $,
one can express the Lyapunov exponent $\lambda$ in terms of $\tilde{T}$. In
FIG. \ref{fig:lambdaT}, $\lambda$ is plotted against $\tilde{T}$ for various
values of $\tilde{Q}$ in the top row. For the case of $\tilde{Q}%
  =0.11<\tilde{Q}_{c}$ in the left column, three black hole solutions (namely
Small BH, Intermediate BH and Large BH) coexist for $\tilde{T}_{1}<\tilde
  {T}<\tilde{T}_{2}$. Moreover, the first-order phase transition between Small
BH and Large BH occurs at $\tilde{T}_{p}$. When $\tilde{T}_{1}<\tilde
  {T}<\tilde{T}_{2}$, $\lambda$ possesses three branches, which exactly match
Small BH, Intermediate BH and Large BH, respectively. For Small BH, $\lambda$
first slightly rises to a maximum and then declines as $\tilde{T}$ increases
toward $\tilde{T}_{2}$. On the other hand, $\lambda$ of Intermediate BH and
Large BH increases and decreases, respectively, with increasing $\tilde{T}$
from $\tilde{T}_{1}$. As expected, $\lambda$ approaches $1$ as $\tilde{T}$
goes to infinity. The right column displays the case of $\tilde{Q}=0.20>Q_{c}%
$, where there is only black hole solution and no phase transition. When
$\tilde{T}$ increases toward infinity, $\lambda$ first increases to a maximum,
then decreases and finally approaches $1$. These observations suggest that the
Lyapunov exponent $\lambda$ can be used to probe phase structure of black holes.

\begin{figure}[t]
  \begin{center}
    \includegraphics[width=0.48\textwidth]{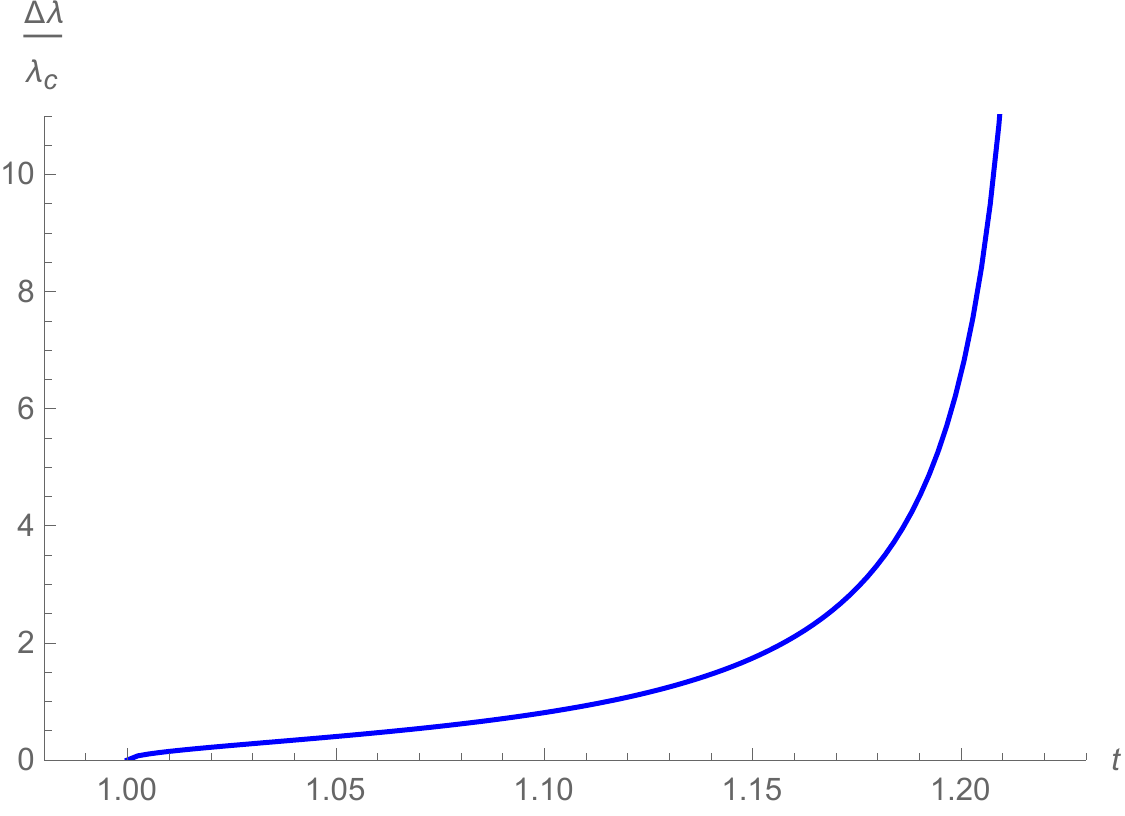}\hspace{1em}
    \includegraphics[width=0.48\textwidth]{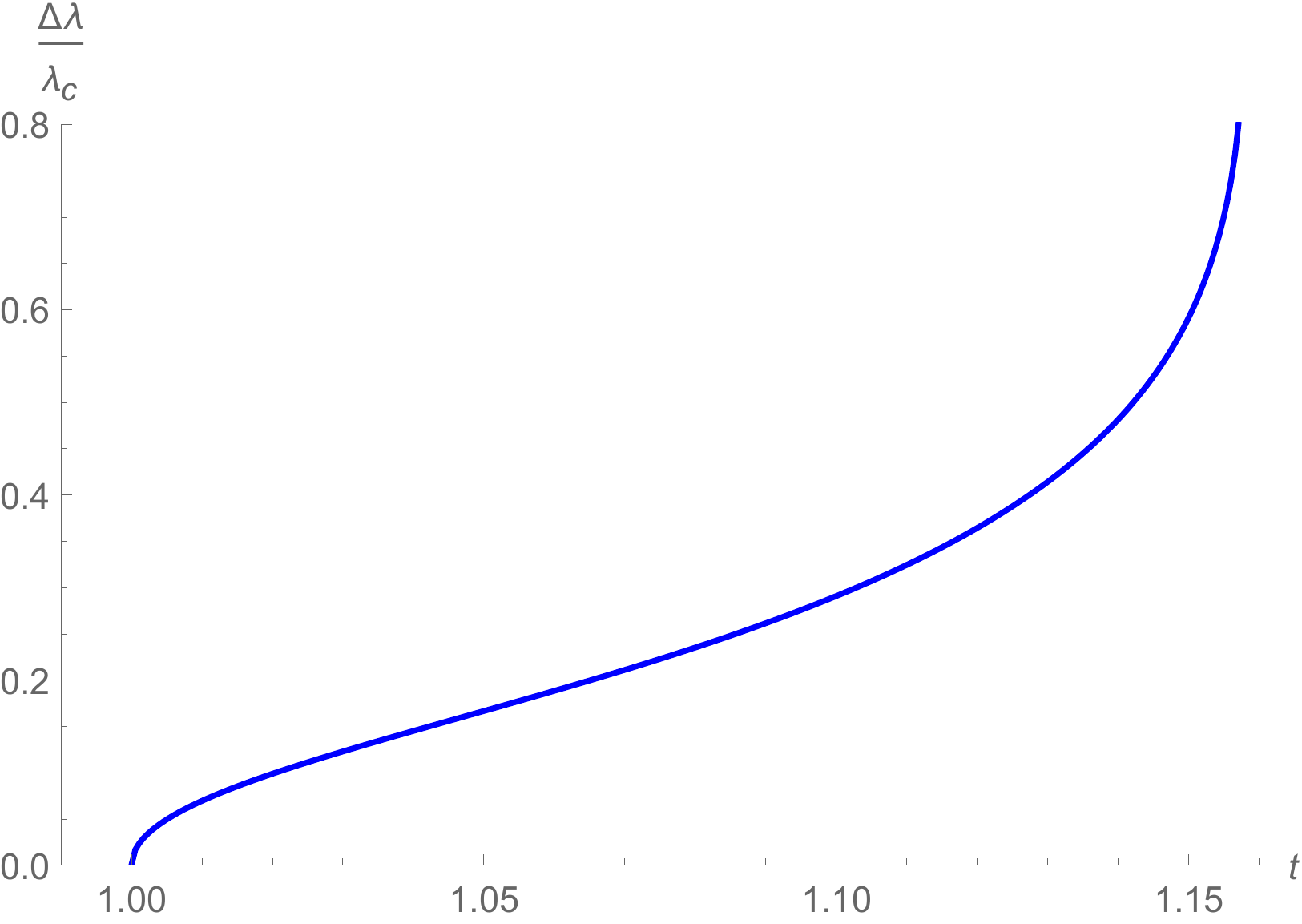}
  \end{center}
  \caption{Rescaled discontinuity in the Lyapunov exponent $\Delta
      \lambda/\lambda_{c}$ during the phase transition as a function of the rescaled
    phase transition temperature $t\equiv\tilde{T}_{p}/\tilde{T}_{c}$ near the
    critical temperature $t=1$. \textbf{Left}: Massless particles on the unstable
    null circular geodesic; \textbf{Right}: Massive particles with $L=20l$ on the
    unstable time-like circular geodesic. The parameter $\Delta\lambda$ is nonzero
    at the first-order phase transition and vanishes at the critical point, which
    indicates that $\Delta\lambda$ plays a role of an order parameter.}%
  \label{dl}%
\end{figure}

Interestingly, the phase transition can be characterized by the discontinuous
change in the Lyapunov exponent, $\Delta\lambda=\lambda_{S}-\lambda_{L}$,
where $\lambda_{S}$ and $\lambda_{L}$ are the Lyapunov exponents of Small BH
and Large BH evaluated at $\tilde{T}=\tilde{T}_{p}$, respectively. Note that,
for the second-order phase transition at the critical point, one has
$\lambda_{S}=\lambda_{L}=\lambda_{c}$, and hence $\Delta\lambda=0$. We display
$\Delta\lambda/\lambda_{c}$ as a function of $t\equiv\tilde{T}_{p}/\tilde
  {T}_{c}$, for which $t=1$ at the critical point, in the left panel of FIG.
\ref{dl}. It shows that, when RN-AdS black holes undergo the first-order phase
transition from Small BH to Large BH, the Lyapunov exponent $\lambda$ jumps
from $\lambda_{S}$ to $\lambda_{L}$ with a nonzero $\Delta\lambda$.
Consequently, $\Delta\lambda$ can be treated as an order parameter. To
investigate the critical behavior of $\Delta\lambda$, we expand $\Delta
  \lambda$ in terms of $t$ near the critical point and obtain%
\begin{equation}
  \frac{\Delta\lambda}{\lambda_{c}}\sim0.700\sqrt{t-1},
\end{equation}
which gives that the critical exponent of $\Delta\lambda$ is $1/2$. Our result
shows that the critical exponent of $\Delta\lambda$ is identical to that of
the order parameter in the van der Waals fluid predicted by the mean field
theory. It is worth emphasizing that the critical exponent of the circular
orbit radius was also found to be $1/2$ for charged AdS black holes
\cite{Wei:2017mwc}.

\subsection{Massive Particles}

\begin{figure}[pt]
  \centering
  \includegraphics[width=10cm]{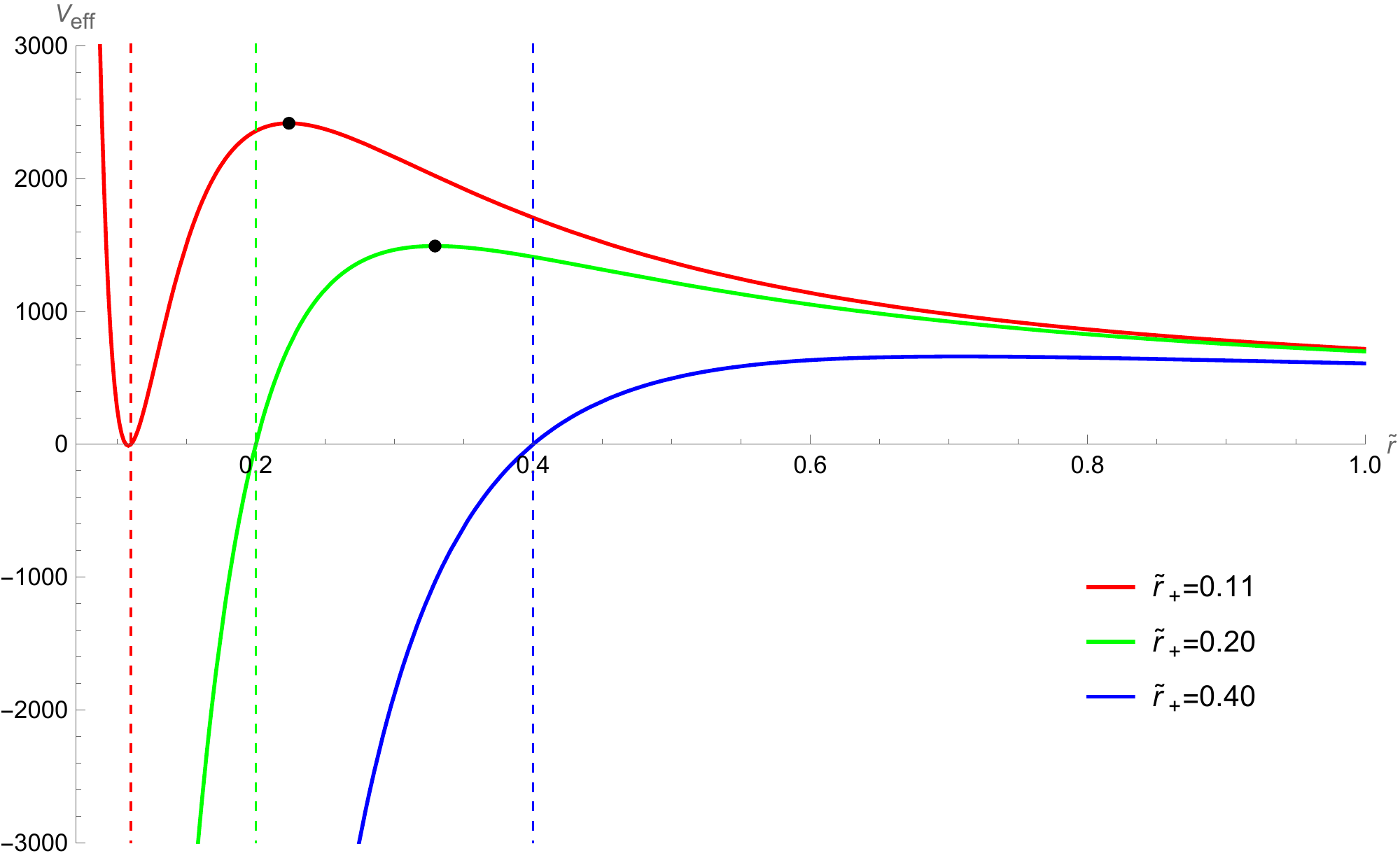}\caption{Effective potential of
    massive particles with the angular momentum $L=20l$ in the RN-AdS black holes
    with $\tilde{Q}=0.11$ for $\tilde{r}_{+}=0.11$, $0.20$ and $0.40$. The
    vertical dashed lines denote the radius of the event horizon. The black dots
    represent the maxima of the effective potentials, corresponding to unstable
    time-like circular geodesics. When $\tilde{r}_{+}=0.40$, the effective
    potential has no maximum.}%
  \label{Veff2}%
\end{figure}

For massive particles, both stable and unstable circular geodesics can exist
in RN-AdS black holes. Since unstable orbits are related to the conjectured
universal upper bound on Lyapunov exponents
\cite{Hashimoto:2016dfz,Zhao:2018wkl}, we here focus on unstable time-like
circular geodesics. Specifically, we consider the Lyapunov exponent of
unstable circular geodesics for massive particles with a given angular
momentum. In FIG. \ref{Veff2}, we plot the effective potential energy of
massive particles with $L=20l$ in RN-AdS black holes with $\tilde{Q}=0.11$ for
various $\tilde{r}_{+}$. It shows that, if $\tilde{r}_{+}$ is too large,
unstable time-like circular geodesics cease to exist.

\begin{figure}[pt]
  \centering
  \includegraphics[width=8cm]{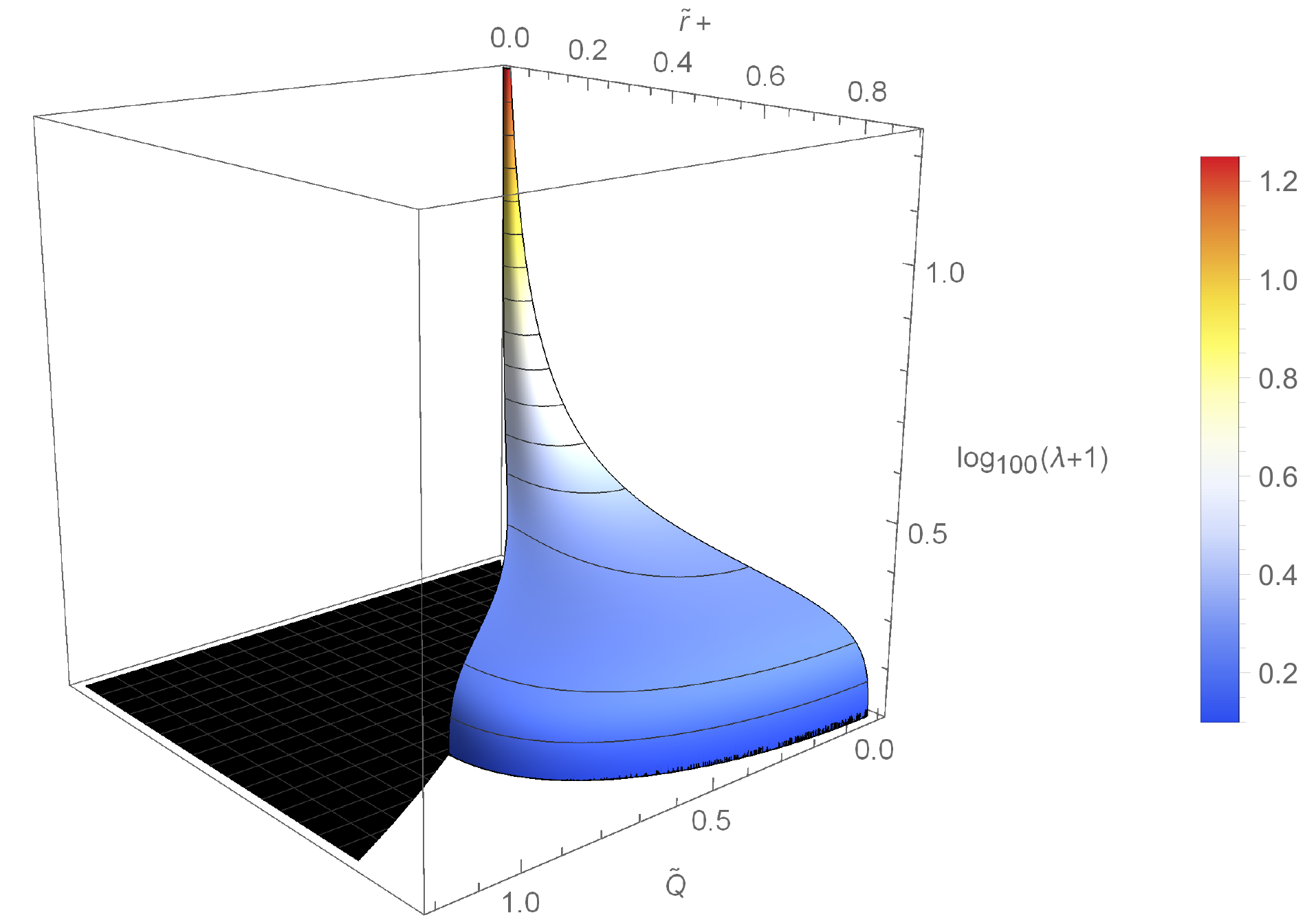} \caption{Three-dimensional plot of
    $\log_{100}{(\lambda+1)}$ as a function of $\tilde{Q}$ and $\tilde{r}_{+}$ for
    massive particles with $L=20l$. There is no black hole solution in the black
    region, and black holes in the white region have no unstable time-like
    circular orbits.}%
  \label{density2}%
\end{figure}

The requirement $V_{\text{eff}}^{\prime}\left(  r\right)  =0$ for a circular
orbit at $r=r_{\text{o}}$ yields
\begin{equation}
  L^{2}=\frac{r_{\text{o}}^{3}f^{\prime}\left(  r_{\text{o}}\right)  }{2f\left(
  r_{\text{o}}\right)  -r_{\text{o}}f^{\prime}\left(  r_{\text{o}}\right)  },
  \label{eq:Ltime}%
\end{equation}
which can be used to express $r_{\text{o}}$ in terms of $L$. The Lyapunov
exponent of the time-like circular orbit is given by \cite{Cardoso:2008bp}
\begin{equation}
  \lambda=\frac{1}{2}\sqrt{\left[  2f(r_{\text{o}})-r_{\text{o}}f^{\prime
  }(r_{\text{o}})\right]  V_{\text{eff}}^{\prime\prime}\left(  r_{\text{o}%
  }\right)  }, \label{eq:lamdaM}%
\end{equation}
which can be rewritten as a function of $L$, $\tilde{Q}$ and $\tilde{r}_{+}$
by using eqn. $\left(  \ref{eq:Ltime}\right)  $. The 3D plot of $\log
  _{100}{({\lambda+}1)}$ as a function of $\tilde{Q}$ and $\tilde{r}_{+}$ is
displayed for $L=20l$ in FIG. \ref{density2}, where black holes in the white
region possess no unstable time-like circular geodesics, and black hole
solutions do not exist in the black region. Note that the Lyapunov exponent
$\lambda$ vanishes on the boundary of the white region.

Plugging $\tilde{r}_{+}(\tilde{T})$ into eqn. $\left(  \ref{eq:lamdaM}\right)
$ leads to $\lambda$ as a function of $\tilde{T}$, which is plotted in the
middle row of FIG. \ref{fig:lambdaT} with various $\tilde{Q}$ and $L=20l$.
Unlike the massless case, there exists a terminate temperature $\tilde{T}_{t}%
$, at which the unstable time-like circular orbit disappears, and $\lambda$
becomes zero. When $\tilde{Q}=0.11<\tilde{Q}_{c}$, $\lambda$ is multivalued
for $\tilde{T}_{1}<\tilde{T}<\tilde{T}_{t}$, for which three black hole
solutions coexist. When $\tilde{Q}=0.20>\tilde{Q}_{c}$, $\lambda$
monotonically decreases and becomes zero at the terminate temperature
$\tilde{T}_{t}$. In addition, the discontinuous change in the Lyapunov
exponent $\Delta\lambda$ is plotted against the temperature $\tilde{T}$ for
massive particles with $L=20l$ in the right panel of FIG. \ref{dl}, which
indicates that $\Delta\lambda$ can serve as an order parameter. Near the
critical temperature, we find%
\begin{equation}
  \frac{\Delta\lambda}{\lambda_{c}}\sim0.731\sqrt{t-1},
\end{equation}
which confirms that the critical exponent of $\Delta\lambda$ is also $1/2$.

\section{Phase Transitions and Lyapunov Exponents of Strings}

\label{sec:string}

\begin{figure}[pt]
  \centering
  \includegraphics[width=4cm]{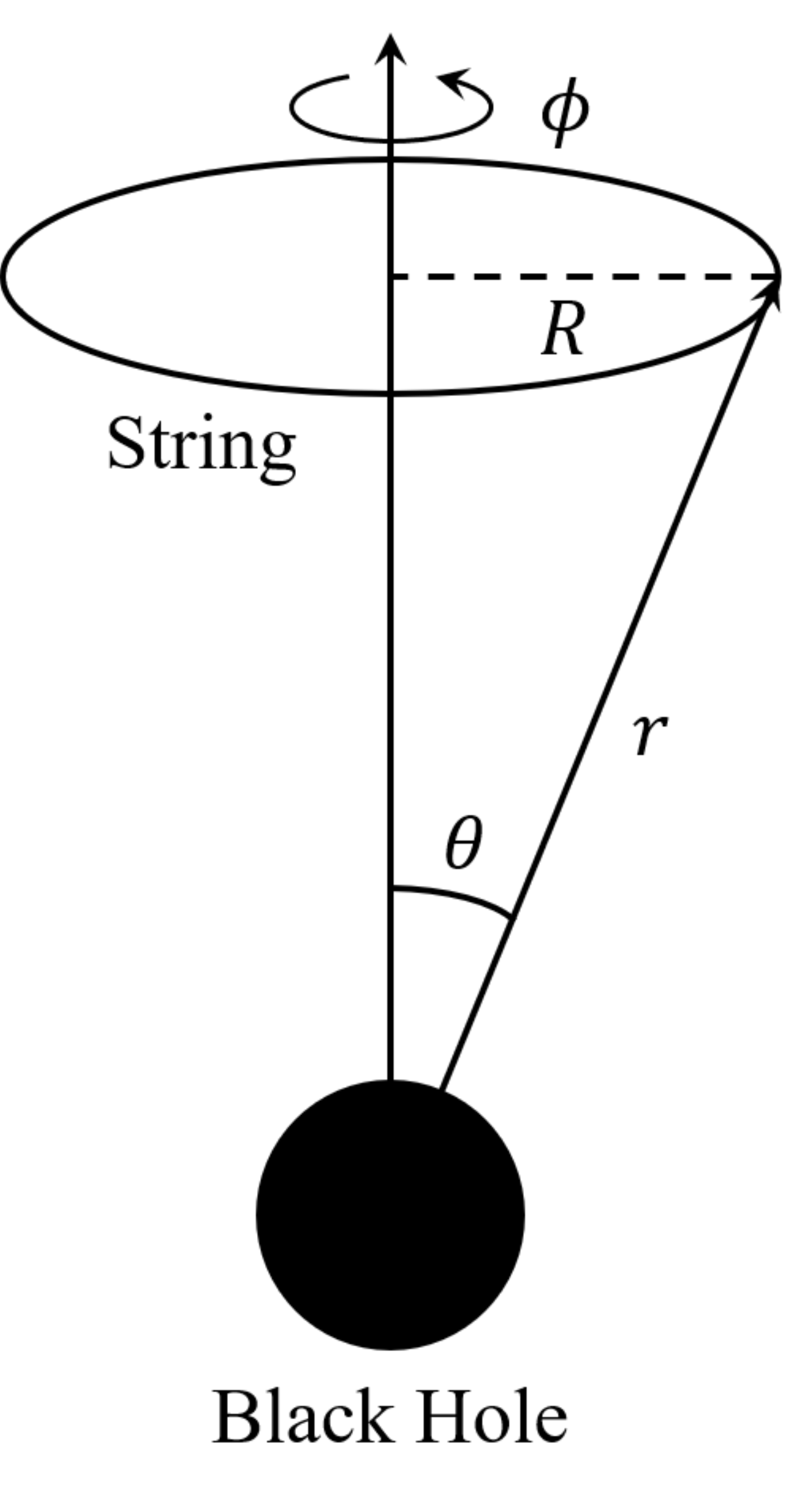} \caption{An oscillating ring string
    moves along the axis of a black hole.}%
  \label{s-b}%
\end{figure}

In contrast to motion of particles in RN-AdS black holes, equations of motion
governing strings are non-integrable, showing chaotic behavior of strings in
RN-AdS black holes \cite{Frolov:1999pj}. Therefore, numerical computations are
often required to obtain Lyapunov exponents of motion of strings. Following
\cite{Basu:2016zkr}, we consider a ring string coaxially moving in the RN-AdS
black hole spacetime, which is illustrated in FIG. \ref{s-b}. The equations of
motion for a string is determined by the Polyakov action,
\begin{equation}
  S_{p}\left(  \gamma,X\right)  =\frac{-1}{2\pi\alpha^{\prime}}\int d\tau
  d\sigma\sqrt{-\gamma}\gamma^{\alpha\beta}\partial_{\alpha}X^{\mu}%
  \partial_{\beta}X^{\nu}G_{\mu\nu},
\end{equation}
where $X^{\mu}$ are the target space coordinates, the indices $\{\alpha
  ,\beta\}=1$ and $2$ correspond to the $(\tau,\sigma)$ coordinates on the
worldsheet of the string, respectively, $\gamma^{\alpha\beta}$ is the
worldsheet metric, and $G_{\mu\nu}$ is the target space metric.

The ring string configuration considered in our paper is described by the
following ansatz for the coordinates of the target space%
\begin{equation}
  t=t(\tau),\ r=r(\tau),\ \theta=\theta(\tau),\ \phi=n\sigma,
\end{equation}
where $n$ is the winding number of the string along the $\phi$ direction, and
$\tau$ is the proper time. For the above ansatz with the conformal gauge
$\gamma^{\alpha\beta}=\eta^{\alpha\beta}$, the Lagrangian for the ring string
in RN-AdS black holes becomes
\begin{equation}
  \mathcal{L}=\frac{-1}{2\pi\alpha^{\prime}}\left[  f\left(  r\right)  \dot
  {t}^{2}-\frac{\dot{r}^{2}}{f\left(  r\right)  }-r^{2}\dot{\theta}^{2}%
  +r^{2}n^{2}\sin^{2}\theta\right]  ,
\end{equation}
where dots and primes denote derivatives with respect to $r$ and $\tau$,
respectively. Using the Legendre transformation, one obtains the Hamiltonian%
\begin{equation}
  \mathcal{H}=\frac{\pi\alpha^{\prime}}{2}\left[  f\left(  r\right)  P_{r}%
      ^{2}+\frac{P_{\theta}^{2}}{r^{2}}-\frac{P_{\theta}^{2}}{f^{2}\left(  r\right)
    }\right]  +\frac{n^{2}r^{2}\sin^{2}\theta}{2\pi\alpha^{\prime}},
\end{equation}
where $P_{t}$, $P_{r}$ and $P_{\theta}$ are the canonical momenta, and the
Hamiltonian satisfies the constraint $\mathcal{H}=0$. The canonical equations
of motion are then given by%
\begin{align}
   & \dot{t}=\frac{-\pi\alpha^{\prime}}{f\left(  r\right)  }P_{t},\nonumber     \\
   & \dot{P_{t}}=0,\nonumber                                                    \\
   & \dot{r}=\pi\alpha^{\prime}f\left(  r\right)  P_{r},\nonumber               \\
   & \dot{P_{r}}=\pi\alpha^{\prime}\left[  -\frac{f^{\prime}\left(  r\right)
      P_{r}^{2}}{2}-\frac{f^{\prime}\left(  r\right)  P_{t}^{2}}{f^{2}\left(
      r\right)  }+\frac{P_{\theta}^{2}}{r^{3}}\right]  -\frac{n^{2}r\sin^{2}\theta
  }{\pi\alpha^{\prime}},                                                        \\
   & \dot{\theta}=\frac{\pi\alpha^{\prime}P_{\theta}}{r^{2}},\nonumber          \\
   & \dot{P_{\theta}}=-\frac{n^{2}r^{2}\sin\theta\cos\theta}{\pi\alpha^{\prime} %
  },\nonumber
\end{align}
which gives that $P_{t}=E$ is the conserved energy. As shown in
\cite{Basu:2016zkr}, there are three different scenarios depending on the
initial conditions of the string and the black hole parameters:

\begin{itemize}
  \item[(1)] The string oscillates back and forth around the black hole.

  \item[(2)] The string oscillates a finite number of times around the black
        hole before being captured by it.

  \item[(3)] The string oscillates a finite number of times around the black
        hole before escaping to infinity.
\end{itemize}

To calculate the Lyapunov exponent of the motion of the string, we evolve two
adjacent trajectories with an initial distance $d_{0}$ in the phase space
spanned by $r$, $\theta$, $P_{r}$ and $P_{\theta}$. When the distance between
the trajectories $d_{t_{i}}$ exceeds the upper threshold at $t=t_{i}$, one
initializes a rescaling of one trajectory back to having the initial distance
$d_{0}$. The maximum Lyapunov exponent is the average of the time-local
Lyapunov exponent
\begin{equation}
  \lambda=\frac{1}{t_{n}-t_{0}}\sum_{i=1}^{n}\ln\left(  \frac{d_{t_{i}}}{d_{0}%
  }\right)  .
\end{equation}
Note that the maximum Lyapunov exponent is of particularly interesting since a
strictly positive maximum Lyapunov exponent can be considered as an indication
of deterministic chaos. Here, we adopt to Verner's \textquotedblleft most
efficient\textquotedblright\ Runge-Kutta $9(8)$ method \cite{verner1996high},
which can achieve high accuracy solving (tolerances like $<10^{-12}$).

In the bottom row of FIG. \ref{fig:lambdaT}, we plot the Lyapunov exponent
$\lambda$ against the temperature $\tilde{T}$ for ring strings of the
scenarios $\left(  1\right)  $ and $\left(  3\right)  $ in RN-AdS black holes
with $\tilde{Q}=0.11$ and $0.20$. Similar to the case of particles, the
Lyapunov exponent $\lambda$ of strings is multivalued when multiple black hole
phases coexist. On the other hand, the Lyapunov exponent $\lambda$ is
single-valued if there exists only one phase. Our results suggest that it is
quite universal to explore phase structure of black holes with Lyapunov exponents.

\section{Conclusions}

\label{sec:conclusion}

In this paper, we calculated Lyapunov exponents of massless particles, massive
particles and ring strings in RN-AdS black holes, and found that the behavior
of the Lyapunov exponents can be employed to explore phase structure of black
holes. In particular, when the black hole charge is less than the critical
charge, the Lyapunov exponents as a function of the temperature demonstrate
three branches, which correspond to three coexisting black hole phases. When
the charge is greater than the critical charge, the Lyapunov exponents are
singled-valued functions of the temperature, which coincides with one black
hole phase. At the first-order phase transition, the discontinuity in the
Lyapunov exponent $\Delta\lambda$ can act as an order parameter to
characterize the black hole phase transition. Remarkably, $\Delta\lambda$ was
shown to have a critical exponent of $1/2$ at the critical point.

Our results support the conjectured relationship between Lyapunov exponents
and phase transition for RN-AdS black holes, which could open a new window to
study thermodynamics of black holes. It will be of great interest if our
analysis can be generalized to more general black hole spacetimes beyond
RN-AdS black holes. More importantly, it is highly desirable to investigate
the relationship between Lyapunov exponents and black hole phase transitions
in the extended phase space, in which the cosmological constant is identified
as a thermodynamic pressure.

\begin{acknowledgments}
  We are grateful to Guangzhou Guo, Xin Jiang and Yiqian Chen for useful
  discussions and valuable comments. This work is supported in part by NSFC
  (Grant Nos. 11747171, 12105191, 11947225 and 11875196), Discipline Talent
  Promotion Program of /Xinglin Scholars (Grant No. QNXZ2018050), Special Talent
  Projects of Chizhou University (Grant No. 2019YJRC001), Anhui Province Natural
  Science Foundation (Grant No. 1808085MA21) and National fund cultivation
  project of Chizhou University (Grant No. CZ2021GP07).
\end{acknowledgments}

\appendix


\begin{thebibliography}{10}

  \bibitem{Hawking:1971tu}
  S.W. Hawking.
  \newblock {Gravitational radiation from colliding black holes}.
  \newblock {\em Phys. Rev. Lett.}, 26:1344--1346, 1971.
  \newblock \href {https://doi.org/10.1103/PhysRevLett.26.1344}
  {\path{doi:10.1103/PhysRevLett.26.1344}}.

  \bibitem{Bekenstein:1972tm}
  Jacob~D Bekenstein.
  \newblock Black holes and the second law.
  \newblock {\em Lett. Nuovo Cim.}, 4(15):737--740, 1972.
  \newblock \href {https://doi.org/10.1007/BF02757029}
  {\path{doi:10.1007/BF02757029}}.

  \bibitem{Bekenstein:1973ur}
  Jacob~D. Bekenstein.
  \newblock {Black holes and entropy}.
  \newblock {\em Phys. Rev. D}, 7:2333--2346, Apr 1973.
  \newblock URL: \url{https://link.aps.org/doi/10.1103/PhysRevD.7.2333}, \href
  {https://doi.org/10.1103/PhysRevD.7.2333}
  {\path{doi:10.1103/PhysRevD.7.2333}}.

  \bibitem{Hawking:1974rv}
  S.W. Hawking.
  \newblock {Black hole explosions}.
  \newblock {\em Nature}, 248:30--31, 1974.
  \newblock \href {https://doi.org/10.1038/248030a0}
  {\path{doi:10.1038/248030a0}}.

  \bibitem{Hawking:1975iha}
  S.~W. Hawking.
  \newblock {Particle Creation by Black Holes}.
  \newblock In {\em {1st Oxford Conference on Quantum Gravity}}, 8 1975.

  \bibitem{Hawking:1982dh}
  S.W. Hawking and Don~N. Page.
  \newblock {Thermodynamics of Black Holes in anti-De Sitter Space}.
  \newblock {\em Commun. Math. Phys.}, 87:577, 1983.
  \newblock \href {https://doi.org/10.1007/BF01208266}
  {\path{doi:10.1007/BF01208266}}.

  \bibitem{Maldacena:1997re}
  Juan~Martin Maldacena.
  \newblock {The Large N limit of superconformal field theories and
    supergravity}.
  \newblock {\em Int. J. Theor. Phys.}, 38:1113--1133, 1999.
  \newblock \href {http://arxiv.org/abs/hep-th/9711200}
  {\path{arXiv:hep-th/9711200}}, \href
  {https://doi.org/10.1023/A:1026654312961}
  {\path{doi:10.1023/A:1026654312961}}.

  \bibitem{Gubser:1998bc}
  S.S. Gubser, Igor~R. Klebanov, and Alexander~M. Polyakov.
  \newblock {Gauge theory correlators from noncritical string theory}.
  \newblock {\em Phys. Lett. B}, 428:105--114, 1998.
  \newblock \href {http://arxiv.org/abs/hep-th/9802109}
  {\path{arXiv:hep-th/9802109}}, \href
  {https://doi.org/10.1016/S0370-2693(98)00377-3}
  {\path{doi:10.1016/S0370-2693(98)00377-3}}.

  \bibitem{Witten:1998qj}
  Edward Witten.
  \newblock {Anti-de Sitter space and holography}.
  \newblock {\em Adv. Theor. Math. Phys.}, 2:253--291, 1998.
  \newblock \href {http://arxiv.org/abs/hep-th/9802150}
  {\path{arXiv:hep-th/9802150}}, \href
  {https://doi.org/10.4310/ATMP.1998.v2.n2.a2}
  {\path{doi:10.4310/ATMP.1998.v2.n2.a2}}.

  \bibitem{Witten:1998zw}
  Edward Witten.
  \newblock {Anti-de Sitter space, thermal phase transition, and confinement in
    gauge theories}.
  \newblock {\em Adv. Theor. Math. Phys.}, 2:505--532, 1998.
  \newblock \href {http://arxiv.org/abs/hep-th/9803131}
  {\path{arXiv:hep-th/9803131}}, \href
  {https://doi.org/10.4310/ATMP.1998.v2.n3.a3}
  {\path{doi:10.4310/ATMP.1998.v2.n3.a3}}.

  \bibitem{Cvetic:1999ne}
  Mirjam Cvetic and Steven~S. Gubser.
  \newblock {Phases of R charged black holes, spinning branes and strongly
    coupled gauge theories}.
  \newblock {\em JHEP}, 04:024, 1999.
  \newblock \href {http://arxiv.org/abs/hep-th/9902195}
  {\path{arXiv:hep-th/9902195}}, \href
  {https://doi.org/10.1088/1126-6708/1999/04/024}
  {\path{doi:10.1088/1126-6708/1999/04/024}}.

  \bibitem{Chamblin:1999tk}
  Andrew Chamblin, Roberto Emparan, Clifford~V. Johnson, and Robert~C. Myers.
  \newblock {Charged AdS black holes and catastrophic holography}.
  \newblock {\em Phys. Rev. D}, 60:064018, Aug 1999.
  \newblock URL: \url{https://link.aps.org/doi/10.1103/PhysRevD.60.064018}, \href
  {http://arxiv.org/abs/hep-th/9902170} {\path{arXiv:hep-th/9902170}}, \href
  {https://doi.org/10.1103/PhysRevD.60.064018}
  {\path{doi:10.1103/PhysRevD.60.064018}}.

  \bibitem{Chamblin:1999hg}
  Andrew Chamblin, Roberto Emparan, Clifford~V. Johnson, and Robert~C. Myers.
  \newblock {Holography, thermodynamics and fluctuations of charged AdS black
    holes}.
  \newblock {\em Phys. Rev. D}, 60:104026, Oct 1999.
  \newblock URL: \url{https://link.aps.org/doi/10.1103/PhysRevD.60.104026}, \href
  {http://arxiv.org/abs/hep-th/9904197} {\path{arXiv:hep-th/9904197}}, \href
  {https://doi.org/10.1103/PhysRevD.60.104026}
  {\path{doi:10.1103/PhysRevD.60.104026}}.

  \bibitem{Caldarelli:1999xj}
  Marco~M. Caldarelli, Guido Cognola, and Dietmar Klemm.
  \newblock {Thermodynamics of Kerr-Newman-AdS black holes and conformal field
    theories}.
  \newblock {\em Class. Quant. Grav.}, 17:399--420, 2000.
  \newblock \href {http://arxiv.org/abs/hep-th/9908022}
  {\path{arXiv:hep-th/9908022}}, \href
  {https://doi.org/10.1088/0264-9381/17/2/310}
  {\path{doi:10.1088/0264-9381/17/2/310}}.

  \bibitem{Cai:2001dz}
  Rong-Gen Cai.
  \newblock {Gauss-Bonnet black holes in AdS spaces}.
  \newblock {\em Phys. Rev. D}, 65:084014, 2002.
  \newblock \href {http://arxiv.org/abs/hep-th/0109133}
  {\path{arXiv:hep-th/0109133}}, \href
  {https://doi.org/10.1103/PhysRevD.65.084014}
  {\path{doi:10.1103/PhysRevD.65.084014}}.

  \bibitem{Cvetic:2001bk}
  Mirjam Cvetic, Shin'ichi Nojiri, and Sergei~D. Odintsov.
  \newblock {Black hole thermodynamics and negative entropy in de Sitter and
    anti-de Sitter Einstein-Gauss-Bonnet gravity}.
  \newblock {\em Nucl. Phys. B}, 628:295--330, 2002.
  \newblock \href {http://arxiv.org/abs/hep-th/0112045}
  {\path{arXiv:hep-th/0112045}}, \href
  {https://doi.org/10.1016/S0550-3213(02)00075-5}
  {\path{doi:10.1016/S0550-3213(02)00075-5}}.

  \bibitem{Nojiri:2001aj}
  Shin'ichi Nojiri and Sergei~D. Odintsov.
  \newblock {Anti-de Sitter black hole thermodynamics in higher derivative
    gravity and new confining deconfining phases in dual CFT}.
  \newblock {\em Phys. Lett. B}, 521:87--95, 2001.
  \newblock [Erratum: Phys.Lett.B 542, 301 (2002)].
  \newblock \href {http://arxiv.org/abs/hep-th/0109122}
  {\path{arXiv:hep-th/0109122}}, \href
  {https://doi.org/10.1016/S0370-2693(01)01186-8}
  {\path{doi:10.1016/S0370-2693(01)01186-8}}.

  \bibitem{Kastor:2009wy}
  Kastor, David and Ray, Sourya and Traschen, Jennie.
  \newblock {Enthalpy and the Mechanics of AdS Black Holes}.
  \newblock {\em Class. Quant. Grav.}, 26:195011, 2009.
  \newblock \href {http://arxiv.org/abs/0904.2765} {\path{arXiv:0904.2765}},
  \href {https://doi.org/10.1088/0264-9381/26/19/195011}
  {\path{doi:10.1088/0264-9381/26/19/195011}}.


  \bibitem{Dolan:2011xt}
  Brian~P. Dolan.
  \newblock {Pressure and volume in the first law of black hole thermodynamics}.
  \newblock {\em Class. Quant. Grav.}, 28:235017, 2011.
  \newblock \href {http://arxiv.org/abs/1106.6260} {\path{arXiv:1106.6260}},
  \href {https://doi.org/10.1088/0264-9381/28/23/235017}
  {\path{doi:10.1088/0264-9381/28/23/235017}}.

  \bibitem{Kubiznak:2012wp}
  David Kubiznak and Robert~B. Mann.
  \newblock {P-V criticality of charged AdS black holes}.
  \newblock {\em JHEP}, 07:033, 2012.
  \newblock \href {http://arxiv.org/abs/1205.0559} {\path{arXiv:1205.0559}},
  \href {https://doi.org/10.1007/JHEP07(2012)033}
  {\path{doi:10.1007/JHEP07(2012)033}}.

  \bibitem{Wei:2012ui}
  Shao-Wen Wei and Yu-Xiao Liu.
  \newblock {Critical phenomena and thermodynamic geometry of charged
    Gauss-Bonnet AdS black holes}.
  \newblock {\em Phys. Rev. D}, 87(4):044014, Feb 2013.
  \newblock URL: \url{https://link.aps.org/doi/10.1103/PhysRevD.87.044014}, \href
  {http://arxiv.org/abs/1209.1707} {\path{arXiv:1209.1707}}, \href
  {https://doi.org/10.1103/PhysRevD.87.044014}
  {\path{doi:10.1103/PhysRevD.87.044014}}.

  \bibitem{Gunasekaran:2012dq}
  Sharmila Gunasekaran, Robert~B. Mann, and David Kubiznak.
  \newblock {Extended phase space thermodynamics for charged and rotating black
    holes and Born-Infeld vacuum polarization}.
  \newblock {\em JHEP}, 11:110, 2012.
  \newblock \href {http://arxiv.org/abs/1208.6251} {\path{arXiv:1208.6251}},
  \href {https://doi.org/10.1007/JHEP11(2012)110}
  {\path{doi:10.1007/JHEP11(2012)110}}.

  \bibitem{Cai:2013qga}
  Rong-Gen Cai, Li-Ming Cao, Li~Li, and Run-Qiu Yang.
  \newblock {P-V criticality in the extended phase space of Gauss-Bonnet black
    holes in AdS space}.
  \newblock {\em JHEP}, 09:005, 2013.
  \newblock \href {http://arxiv.org/abs/1306.6233} {\path{arXiv:1306.6233}},
  \href {https://doi.org/10.1007/JHEP09(2013)005}
  {\path{doi:10.1007/JHEP09(2013)005}}.

  \bibitem{Altamirano:2013ane}
  Natacha Altamirano, David Kubiznak, and Robert~B. Mann.
  \newblock {Reentrant phase transitions in rotating anti\textendash{}de Sitter
    black holes}.
  \newblock {\em Phys. Rev. D}, 88(10):101502, 2013.
  \newblock \href {http://arxiv.org/abs/1306.5756} {\path{arXiv:1306.5756}},
  \href {https://doi.org/10.1103/PhysRevD.88.101502}
  {\path{doi:10.1103/PhysRevD.88.101502}}.

  \bibitem{Altamirano:2013uqa}
  Natacha Altamirano, David Kubiz\v~nák, Robert~B. Mann, and Zeinab
  Sherkatghanad.
  \newblock {Kerr-AdS analogue of triple point and solid/liquid/gas phase
    transition}.
  \newblock {\em Class. Quant. Grav.}, 31:042001, 2014.
  \newblock \href {http://arxiv.org/abs/1308.2672} {\path{arXiv:1308.2672}},
  \href {https://doi.org/10.1088/0264-9381/31/4/042001}
  {\path{doi:10.1088/0264-9381/31/4/042001}}.

  \bibitem{Xu:2014kwa}
  Wei Xu and Liu Zhao.
  \newblock {Critical phenomena of static charged AdS black holes in conformal
    gravity}.
  \newblock {\em Phys. Lett. B}, 736:214--220, 2014.
  \newblock \href {http://arxiv.org/abs/1405.7665} {\path{arXiv:1405.7665}},
  \href {https://doi.org/10.1016/j.physletb.2014.07.019}
  {\path{doi:10.1016/j.physletb.2014.07.019}}.

  \bibitem{Frassino:2014pha}
  Antonia~M. Frassino, David Kubiznak, Robert~B. Mann, and Fil Simovic.
  \newblock {Multiple Reentrant Phase Transitions and Triple Points in Lovelock
    Thermodynamics}.
  \newblock {\em JHEP}, 09:080, 2014.
  \newblock \href {http://arxiv.org/abs/1406.7015} {\path{arXiv:1406.7015}},
  \href {https://doi.org/10.1007/JHEP09(2014)080}
  {\path{doi:10.1007/JHEP09(2014)080}}.

  \bibitem{Dehghani:2014caa}
  M.~H. Dehghani, S.~Kamrani, and A.~Sheykhi.
  \newblock {$P-V$ criticality of charged dilatonic black holes}.
  \newblock {\em Phys. Rev. D}, 90(10):104020, 2014.
  \newblock \href {http://arxiv.org/abs/1505.02386} {\path{arXiv:1505.02386}},
  \href {https://doi.org/10.1103/PhysRevD.90.104020}
  {\path{doi:10.1103/PhysRevD.90.104020}}.

  \bibitem{Wei:2014hba}
  Shao-Wen Wei and Yu-Xiao Liu.
  \newblock {Triple points and phase diagrams in the extended phase space of
    charged Gauss-Bonnet black holes in AdS space}.
  \newblock {\em Phys. Rev. D}, 90(4):044057, 2014.
  \newblock \href {http://arxiv.org/abs/1402.2837} {\path{arXiv:1402.2837}},
  \href {https://doi.org/10.1103/PhysRevD.90.044057}
  {\path{doi:10.1103/PhysRevD.90.044057}}.

  \bibitem{Dolan:2014vba}
  Brian~P. Dolan, Anna Kostouki, David Kubiznak, and Robert~B. Mann.
  \newblock {Isolated critical point from Lovelock gravity}.
  \newblock {\em Class. Quant. Grav.}, 31(24):242001, 2014.
  \newblock \href {http://arxiv.org/abs/1407.4783} {\path{arXiv:1407.4783}},
  \href {https://doi.org/10.1088/0264-9381/31/24/242001}
  {\path{doi:10.1088/0264-9381/31/24/242001}}.

  \bibitem{Hennigar:2015esa}
  Robie~A. Hennigar, Wilson~G. Brenna, and Robert~B. Mann.
  \newblock {$P-v$ criticality in quasitopological gravity}.
  \newblock {\em JHEP}, 07:077, 2015.
  \newblock \href {http://arxiv.org/abs/1505.05517} {\path{arXiv:1505.05517}},
  \href {https://doi.org/10.1007/JHEP07(2015)077}
  {\path{doi:10.1007/JHEP07(2015)077}}.

  \bibitem{Caceres:2015vsa}
  Elena Caceres, Phuc~H. Nguyen, and Juan~F. Pedraza.
  \newblock {Holographic entanglement entropy and the extended phase structure of
    STU black holes}.
  \newblock {\em JHEP}, 09:184, 2015.
  \newblock \href {http://arxiv.org/abs/1507.06069} {\path{arXiv:1507.06069}},
  \href {https://doi.org/10.1007/JHEP09(2015)184}
  {\path{doi:10.1007/JHEP09(2015)184}}.

  \bibitem{Wei:2015ana}
  Shao-Wen Wei, Peng Cheng, and Yu-Xiao Liu.
  \newblock {Analytical and exact critical phenomena of $d$-dimensional singly
    spinning Kerr-AdS black holes}.
  \newblock {\em Phys. Rev. D}, 93(8):084015, 2016.
  \newblock \href {http://arxiv.org/abs/1510.00085} {\path{arXiv:1510.00085}},
  \href {https://doi.org/10.1103/PhysRevD.93.084015}
  {\path{doi:10.1103/PhysRevD.93.084015}}.

  \bibitem{Chakraborty:2015hna}
  Sumanta Chakraborty and T.~Padmanabhan.
  \newblock {Thermodynamical interpretation of the geometrical variables
    associated with null surfaces}.
  \newblock {\em Phys. Rev. D}, 92(10):104011, 2015.
  \newblock \href {http://arxiv.org/abs/1508.04060} {\path{arXiv:1508.04060}},
  \href {https://doi.org/10.1103/PhysRevD.92.104011}
  {\path{doi:10.1103/PhysRevD.92.104011}}.

  \bibitem{Hendi:2016yof}
  Seyed~Hossein Hendi, Gu-Qiang Li, Jie-Xiong Mo, Shahram Panahiyan, and Behzad
  Eslam~Panah.
  \newblock {New perspective for black hole thermodynamics in
    Gauss--Bonnet--Born--Infeld massive gravity}.
  \newblock {\em Eur. Phys. J. C}, 76(10):571, 2016.
  \newblock \href {http://arxiv.org/abs/1608.03148} {\path{arXiv:1608.03148}},
  \href {https://doi.org/10.1140/epjc/s10052-016-4410-4}
  {\path{doi:10.1140/epjc/s10052-016-4410-4}}.

  \bibitem{Hennigar:2016xwd}
  Robie~A. Hennigar, Robert~B. Mann, and Erickson Tjoa.
  \newblock {Superfluid Black Holes}.
  \newblock {\em Phys. Rev. Lett.}, 118(2):021301, 2017.
  \newblock \href {http://arxiv.org/abs/1609.02564} {\path{arXiv:1609.02564}},
  \href {https://doi.org/10.1103/PhysRevLett.118.021301}
  {\path{doi:10.1103/PhysRevLett.118.021301}}.

  \bibitem{Momeni:2016qfv}
  Davood Momeni, Mir Faizal, Kairat Myrzakulov, and Ratbay Myrzakulov.
  \newblock {Fidelity Susceptibility as Holographic PV-Criticality}.
  \newblock {\em Phys. Lett. B}, 765:154--158, 2017.
  \newblock \href {http://arxiv.org/abs/1604.06909} {\path{arXiv:1604.06909}},
  \href {https://doi.org/10.1016/j.physletb.2016.12.006}
  {\path{doi:10.1016/j.physletb.2016.12.006}}.

  \bibitem{Hendi:2017fxp}
  S.H. Hendi, R.B. Mann, S.~Panahiyan, and B.~Eslam~Panah.
  \newblock {Van der Waals like behavior of topological AdS black holes in
    massive gravity}.
  \newblock {\em Phys. Rev. D}, 95(2):021501, 2017.
  \newblock \href {http://arxiv.org/abs/1702.00432} {\path{arXiv:1702.00432}},
  \href {https://doi.org/10.1103/PhysRevD.95.021501}
  {\path{doi:10.1103/PhysRevD.95.021501}}.

  \bibitem{Lemos:2018cfd}
  José~P.S. Lemos and Oleg~B. Zaslavskii.
  \newblock {Black hole thermodynamics with the cosmological constant as
    independent variable: Bridge between the enthalpy and the Euclidean path
    integral approaches}.
  \newblock {\em Phys. Lett. B}, 786:296--299, 2018.
  \newblock \href {http://arxiv.org/abs/1806.07910} {\path{arXiv:1806.07910}},
  \href {https://doi.org/10.1016/j.physletb.2018.08.075}
  {\path{doi:10.1016/j.physletb.2018.08.075}}.

  \bibitem{Pedraza:2018eey}
  Juan~F. Pedraza, Watse Sybesma, and Manus~R. Visser.
  \newblock {Hyperscaling violating black holes with spherical and hyperbolic
    horizons}.
  \newblock {\em Class. Quant. Grav.}, 36(5):054002, 2019.
  \newblock \href {http://arxiv.org/abs/1807.09770} {\path{arXiv:1807.09770}},
  \href {https://doi.org/10.1088/1361-6382/ab0094}
  {\path{doi:10.1088/1361-6382/ab0094}}.

  \bibitem{Wang:2018xdz}
  Peng Wang, Houwen Wu, and Haitang Yang.
  \newblock {Thermodynamics and Phase Transitions of Nonlinear Electrodynamics
    Black Holes in an Extended Phase Space}.
  \newblock {\em JCAP}, 04(04):052, 2019.
  \newblock \href {http://arxiv.org/abs/1808.04506} {\path{arXiv:1808.04506}},
  \href {https://doi.org/10.1088/1475-7516/2019/04/052}
  {\path{doi:10.1088/1475-7516/2019/04/052}}.

  \bibitem{Wei:2020poh}
  Shao-Wen Wei and Yu-Xiao Liu.
  \newblock {Extended thermodynamics and microstructures of four-dimensional
    charged Gauss-Bonnet black hole in AdS space}.
  \newblock {\em Phys. Rev. D}, 101(10):104018, 2020.
  \newblock \href {http://arxiv.org/abs/2003.14275} {\path{arXiv:2003.14275}},
  \href {https://doi.org/10.1103/PhysRevD.101.104018}
  {\path{doi:10.1103/PhysRevD.101.104018}}.

  \bibitem{Ruppeiner:2012uc}
  George Ruppeiner.
  \newblock {Thermodynamic curvature: pure fluids to black holes}.
  \newblock {\em J. Phys. Conf. Ser.}, 410:012138, 2013.
  \newblock \href {http://arxiv.org/abs/1210.2011} {\path{arXiv:1210.2011}},
  \href {https://doi.org/10.1088/1742-6596/410/1/012138}
  {\path{doi:10.1088/1742-6596/410/1/012138}}.

  \bibitem{Miao:2017cyt}
  Yan-Gang Miao and Zhen-Ming Xu.
  \newblock {Microscopic structures and thermal stability of black holes
    conformally coupled to scalar fields in five dimensions}.
  \newblock {\em Nucl. Phys. B}, 942:205--220, 2019.
  \newblock \href {http://arxiv.org/abs/1711.01757} {\path{arXiv:1711.01757}},
  \href {https://doi.org/10.1016/j.nuclphysb.2019.03.015}
  {\path{doi:10.1016/j.nuclphysb.2019.03.015}}.

  \bibitem{Guo:2019oad}
  Xiong-Ying Guo, Huai-Fan Li, Li-Chun Zhang, and Ren Zhao.
  \newblock {Microstructure and continuous phase transition of a
    Reissner-Nordstrom-AdS black hole}.
  \newblock {\em Phys. Rev. D}, 100(6):064036, 2019.
  \newblock \href {http://arxiv.org/abs/1901.04703} {\path{arXiv:1901.04703}},
  \href {https://doi.org/10.1103/PhysRevD.100.064036}
  {\path{doi:10.1103/PhysRevD.100.064036}}.

  \bibitem{Wei:2019yvs}
  Shao-Wen Wei, Yu-Xiao Liu, and Robert~B. Mann.
  \newblock {Ruppeiner Geometry, Phase Transitions, and the Microstructure of
    Charged AdS Black Holes}.
  \newblock {\em Phys. Rev. D}, 100(12):124033, 2019.
  \newblock \href {http://arxiv.org/abs/1909.03887} {\path{arXiv:1909.03887}},
  \href {https://doi.org/10.1103/PhysRevD.100.124033}
  {\path{doi:10.1103/PhysRevD.100.124033}}.

  \bibitem{Wang:2019cax}
  Peng Wang, Houwen Wu, and Haitang Yang.
  \newblock {Thermodynamic Geometry of AdS Black Holes and Black Holes in a
    Cavity}.
  \newblock {\em Eur. Phys. J. C}, 80(3):216, 2020.
  \newblock \href {http://arxiv.org/abs/1910.07874} {\path{arXiv:1910.07874}},
  \href {https://doi.org/10.1140/epjc/s10052-020-7776-2}
  {\path{doi:10.1140/epjc/s10052-020-7776-2}}.

  \bibitem{Yerra:2020oph}
  Yerra, Pavan Kumar and Bhamidipati, Chandrasekhar.
  \newblock {Ruppeiner Geometry, Phase Transitions and Microstructures of Black Holes in Massive Gravity}.
  \newblock {\em Int. J. Mod. Phys. A}, 35(22):2050120, 2020.
  \newblock \href {http://arxiv.org/abs/2006.07775} {\path{arXiv:2006.07775}},
  \href {https://doi.org/10.1142/S0217751X20501201}
  {\path{doi:10.1142/S0217751X20501201}}.

  \bibitem{Yerra:2021hnh}
  Yerra, Pavan Kumar and Bhamidipati, Chandrasekhar.
  \newblock {Novel relations in massive gravity at Hawking-Page transition}.
  \newblock {\em Phys. Rev. D}, 104(10):104049, 2021.
  \newblock \href {http://arxiv.org/abs/2107.04504} {\path{arXiv:2107.04504}},
  \href {https://doi.org/10.1103/PhysRevD.104.104049}
  {\path{doi:10.1103/PhysRevD.104.104049}}.

  \bibitem{Wei:2015iwa}
  Shao-Wen Wei and Yu-Xiao Liu.
  \newblock {Insight into the Microscopic Structure of an AdS Black Hole from a
    Thermodynamical Phase Transition}.
  \newblock {\em Phys. Rev. Lett.}, 115(11):111302, 2015.
  \newblock [Erratum: Phys.Rev.Lett. 116, 169903 (2016)].
  \newblock \href {http://arxiv.org/abs/1502.00386} {\path{arXiv:1502.00386}},
  \href {https://doi.org/10.1103/PhysRevLett.115.111302}
  {\path{doi:10.1103/PhysRevLett.115.111302}}.

  \bibitem{Wei:2019uqg}
  Shao-Wen Wei, Yu-Xiao Liu, and Robert~B. Mann.
  \newblock {Repulsive Interactions and Universal Properties of Charged Anti--de
    Sitter Black Hole Microstructures}.
  \newblock {\em Phys. Rev. Lett.}, 123(7):071103, 2019.
  \newblock \href {http://arxiv.org/abs/1906.10840} {\path{arXiv:1906.10840}},
  \href {https://doi.org/10.1103/PhysRevLett.123.071103}
  {\path{doi:10.1103/PhysRevLett.123.071103}}.

  \bibitem{Liu:2014gvf}
  Yunqi Liu, De-Cheng Zou, and Bin Wang.
  \newblock {Signature of the Van der Waals like small-large charged AdS black
    hole phase transition in quasinormal modes}.
  \newblock {\em JHEP}, 09:179, 2014.
  \newblock \href {http://arxiv.org/abs/1405.2644} {\path{arXiv:1405.2644}},
  \href {https://doi.org/10.1007/JHEP09(2014)179}
  {\path{doi:10.1007/JHEP09(2014)179}}.

  \bibitem{Mahapatra:2016dae}
  Subhash Mahapatra.
  \newblock {Thermodynamics, Phase Transition and Quasinormal modes with Weyl
    corrections}.
  \newblock {\em JHEP}, 04:142, 2016.
  \newblock \href {http://arxiv.org/abs/1602.03007} {\path{arXiv:1602.03007}},
  \href {https://doi.org/10.1007/JHEP04(2016)142}
  {\path{doi:10.1007/JHEP04(2016)142}}.

  \bibitem{Chabab:2016cem}
  M.~Chabab, H.~El~Moumni, S.~Iraoui, and K.~Masmar.
  \newblock {Behavior of quasinormal modes and high dimension RN\textendash{}AdS
    black hole phase transition}.
  \newblock {\em Eur. Phys. J. C}, 76(12):676, 2016.
  \newblock \href {http://arxiv.org/abs/1606.08524} {\path{arXiv:1606.08524}},
  \href {https://doi.org/10.1140/epjc/s10052-016-4518-6}
  {\path{doi:10.1140/epjc/s10052-016-4518-6}}.

  \bibitem{Zou:2017juz}
  De-Cheng Zou, Yunqi Liu, and Rui-Hong Yue.
  \newblock {Behavior of quasinormal modes and Van der Waals-like phase
    transition of charged AdS black holes in massive gravity}.
  \newblock {\em Eur. Phys. J. C}, 77(6):365, 2017.
  \newblock \href {http://arxiv.org/abs/1702.08118} {\path{arXiv:1702.08118}},
  \href {https://doi.org/10.1140/epjc/s10052-017-4937-z}
  {\path{doi:10.1140/epjc/s10052-017-4937-z}}.

  \bibitem{Zhang:2020khz}
  Ming Zhang, Chao-Ming Zhang, De-Cheng Zou, and Rui-Hong Yue.
  \newblock {Phase transition and Quasinormal modes for Charged black holes in 4D
    Einstein-Gauss-Bonnet gravity}.
  \newblock {\em Chin. Phys. C}, 45(4):045105, 2021.
  \newblock \href {http://arxiv.org/abs/2009.03096} {\path{arXiv:2009.03096}},
  \href {https://doi.org/10.1088/1674-1137/abe19a}
  {\path{doi:10.1088/1674-1137/abe19a}}.

  \bibitem{Wei:2017mwc}
  Shao-Wen Wei and Yu-Xiao Liu.
  \newblock {Photon orbits and thermodynamic phase transition of $d$-dimensional
    charged AdS black holes}.
  \newblock {\em Phys. Rev. D}, 97(10):104027, 2018.
  \newblock \href {http://arxiv.org/abs/1711.01522} {\path{arXiv:1711.01522}},
  \href {https://doi.org/10.1103/PhysRevD.97.104027}
  {\path{doi:10.1103/PhysRevD.97.104027}}.

  \bibitem{Wei:2018aqm}
  Shao-Wen Wei, Yu-Xiao Liu, and Yong-Qiang Wang.
  \newblock {Probing the relationship between the null geodesics and
    thermodynamic phase transition for rotating Kerr-AdS black holes}.
  \newblock {\em Phys. Rev. D}, 99(4):044013, 2019.
  \newblock \href {http://arxiv.org/abs/1807.03455} {\path{arXiv:1807.03455}},
  \href {https://doi.org/10.1103/PhysRevD.99.044013}
  {\path{doi:10.1103/PhysRevD.99.044013}}.

  \bibitem{Zhang:2019tzi}
  Ming Zhang, Shan-Zhong Han, Jie Jiang, and Wen-Biao Liu.
  \newblock {Circular orbit of a test particle and phase transition of a black
    hole}.
  \newblock {\em Phys. Rev. D}, 99(6):065016, 2019.
  \newblock \href {http://arxiv.org/abs/1903.08293} {\path{arXiv:1903.08293}},
  \href {https://doi.org/10.1103/PhysRevD.99.065016}
  {\path{doi:10.1103/PhysRevD.99.065016}}.

  \bibitem{Zhang:2019glo}
  Ming Zhang and Minyong Guo.
  \newblock {Can shadows reflect phase structures of black holes?}
  \newblock {\em Eur. Phys. J. C}, 80(8):790, 2020.
  \newblock \href {http://arxiv.org/abs/1909.07033} {\path{arXiv:1909.07033}},
  \href {https://doi.org/10.1140/epjc/s10052-020-8389-5}
  {\path{doi:10.1140/epjc/s10052-020-8389-5}}.

  \bibitem{Belhaj:2020nqy}
  A.~Belhaj, L.~Chakhchi, H.~El~Moumni, J.~Khalloufi, and K.~Masmar.
  \newblock {Thermal Image and Phase Transitions of Charged AdS Black Holes using
    Shadow Analysis}.
  \newblock {\em Int. J. Mod. Phys. A}, 35(27):2050170, 2020.
  \newblock \href {http://arxiv.org/abs/2005.05893} {\path{arXiv:2005.05893}},
  \href {https://doi.org/10.1142/S0217751X20501705}
  {\path{doi:10.1142/S0217751X20501705}}.

  \bibitem{lyapunov1992general}
  Aleksandr~Mikhailovich Lyapunov.
  \newblock The general problem of the stability of motion.
  \newblock {\em International journal of control}, 55(3):531--534, 1992.
  \newblock \href
  {http://arxiv.org/abs/https://doi.org/10.1080/00207179208934253}
  {\path{arXiv:https://doi.org/10.1080/00207179208934253}}, \href
  {https://doi.org/10.1080/00207179208934253}
  {\path{doi:10.1080/00207179208934253}}.

  \bibitem{Sota:1995ms}
  Yasuhide Sota, Shingo Suzuki, and Kei-ichi Maeda.
  \newblock {Chaos in static axisymmetric space-times. 1: Vacuum case}.
  \newblock {\em Class. Quant. Grav.}, 13:1241--1260, 1996.
  \newblock \href {http://arxiv.org/abs/gr-qc/9505036}
  {\path{arXiv:gr-qc/9505036}}, \href
  {https://doi.org/10.1088/0264-9381/13/5/034}
  {\path{doi:10.1088/0264-9381/13/5/034}}.

  \bibitem{Sota:1996cv}
  Yasuhide Sota, Shingo Suzuki, and Kei-ichi Maeda.
  \newblock {Chaos in static axisymmetric space-times. 2. Nonvacuum case}.
  \newblock 10 1996.
  \newblock \href {http://arxiv.org/abs/gr-qc/9610065}
  {\path{arXiv:gr-qc/9610065}}.

  \bibitem{Kan:2021blg}
  N.~Kan and B.~Gwak.
  \newblock{Bound on the Lyapunov exponent in Kerr-Newman black holes via a charged particle}.
  \newblock {\em  Phys. Rev. D }, 105(2):026006, 2022.
  \newblock \href {http://arxiv.org/abs/2109.07341}
  {\path{arXiv:2109.07341}}, \href
  {https://doi.org/10.1103/PhysRevD.105.026006}
  {\path{doi:=10.1103/PhysRevD.105.026006}}.

  \bibitem{Gwak:2022xje}
  Gwak, Bogeun and Kan, Naoto and Lee, Bum-Hoon and Lee, Hocheol.
  \newblock{Violation of bound on chaos for charged probe in Kerr-Newman-AdS black hole}.
  \newblock \href {http://arxiv.org/abs/2203.07298}
  {\path{arXiv:2203.07298}}.

  \bibitem{Hanan:2006uf}
  William Hanan and Eugen Radu.
  \newblock {Chaotic motion in multi-black hole spacetimes and holographic
    screens}.
  \newblock {\em Mod. Phys. Lett. A}, 22:399--406, 2007.
  \newblock \href {http://arxiv.org/abs/gr-qc/0610119}
  {\path{arXiv:gr-qc/0610119}}, \href
  {https://doi.org/10.1142/S0217732307022815}
  {\path{doi:10.1142/S0217732307022815}}.

  \bibitem{Gair:2007kr}
  Jonathan~R. Gair, Chao Li, and Ilya Mandel.
  \newblock {Observable Properties of Orbits in Exact Bumpy Spacetimes}.
  \newblock {\em Phys. Rev. D}, 77:024035, 2008.
  \newblock \href {http://arxiv.org/abs/0708.0628} {\path{arXiv:0708.0628}},
  \href {https://doi.org/10.1103/PhysRevD.77.024035}
  {\path{doi:10.1103/PhysRevD.77.024035}}.

  \bibitem{Zahrani:2013up}
  A.~M.~Al Zahrani, Valeri~P. Frolov, and Andrey~A. Shoom.
  \newblock {Critical escape velocity for a charged particle moving around a
    weakly magnetized Schwarzschild black hole}.
  \newblock {\em Phys. Rev. D}, 87(8):084043, 2013.
  \newblock \href {http://arxiv.org/abs/1301.4633} {\path{arXiv:1301.4633}},
  \href {https://doi.org/10.1103/PhysRevD.87.084043}
  {\path{doi:10.1103/PhysRevD.87.084043}}.

  \bibitem{Polcar:2019kwu}
  L.~Polcar and O.~Semer\'ak.
  \newblock {Free motion around black holes with discs or rings: Between
    integrability and chaos. VI. The Melnikov method}.
  \newblock {\em Phys. Rev. D}, 100(10):103013, 2019.
  \newblock \href {http://arxiv.org/abs/1911.09790} {\path{arXiv:1911.09790}},
  \href {https://doi.org/10.1103/PhysRevD.100.103013}
  {\path{doi:10.1103/PhysRevD.100.103013}}.

  \bibitem{Wang:2016wcj}
  Mingzhi Wang, Songbai Chen, and Jiliang Jing.
  \newblock {Chaos in the motion of a test scalar particle coupling to the
    Einstein tensor in Schwarzschild\textendash{}Melvin black hole spacetime}.
  \newblock {\em Eur. Phys. J. C}, 77(4):208, 2017.
  \newblock \href {http://arxiv.org/abs/1605.09506} {\path{arXiv:1605.09506}},
  \href {https://doi.org/10.1140/epjc/s10052-017-4792-y}
  {\path{doi:10.1140/epjc/s10052-017-4792-y}}.

  \bibitem{Chen:2016tmr}
  Songbai Chen, Mingzhi Wang, and Jiliang Jing.
  \newblock {Chaotic motion of particles in the accelerating and rotating black
    holes spacetime}.
  \newblock {\em JHEP}, 09:082, 2016.
  \newblock \href {http://arxiv.org/abs/1604.02785} {\path{arXiv:1604.02785}},
  \href {https://doi.org/10.1007/JHEP09(2016)082}
  {\path{doi:10.1007/JHEP09(2016)082}}.

  \bibitem{Wang:2018eui}
  Mingzhi Wang, Songbai Chen, and Jiliang Jing.
  \newblock {Chaotic shadow of a non-Kerr rotating compact object with quadrupole
    mass moment}.
  \newblock {\em Phys. Rev. D}, 98(10):104040, 2018.
  \newblock \href {http://arxiv.org/abs/1801.02118} {\path{arXiv:1801.02118}},
  \href {https://doi.org/10.1103/PhysRevD.98.104040}
  {\path{doi:10.1103/PhysRevD.98.104040}}.

  \bibitem{Lu:2018mpr}
  Fenghua Lu, Jun Tao, and Peng Wang.
  \newblock {Minimal Length Effects on Chaotic Motion of Particles around Black
    Hole Horizon}.
  \newblock {\em JCAP}, 12:036, 2018.
  \newblock \href {http://arxiv.org/abs/1811.02140} {\path{arXiv:1811.02140}},
  \href {https://doi.org/10.1088/1475-7516/2018/12/036}
  {\path{doi:10.1088/1475-7516/2018/12/036}}.

  \bibitem{Guo:2020xnf}
  Xiaobo Guo, Kangkai Liang, Benrong Mu, Peng Wang, and Mingtao Yang.
  \newblock {Chaotic Motion around a Black Hole under Minimal Length Effects}.
  \newblock {\em Eur. Phys. J. C}, 80(8):745, 2020.
  \newblock \href {http://arxiv.org/abs/2002.05894} {\path{arXiv:2002.05894}},
  \href {https://doi.org/10.1140/epjc/s10052-020-8335-6}
  {\path{doi:10.1140/epjc/s10052-020-8335-6}}.

  \bibitem{Hashimoto:2016dfz}
  Koji Hashimoto and Norihiro Tanahashi.
  \newblock {Universality in Chaos of Particle Motion near Black Hole Horizon}.
  \newblock {\em Phys. Rev. D}, 95(2):024007, 2017.
  \newblock \href {http://arxiv.org/abs/1610.06070} {\path{arXiv:1610.06070}},
  \href {https://doi.org/10.1103/PhysRevD.95.024007}
  {\path{doi:10.1103/PhysRevD.95.024007}}.

  \bibitem{Dalui:2018qqv}
  Surojit Dalui, Bibhas~Ranjan Majhi, and Pankaj Mishra.
  \newblock {Presence of horizon makes particle motion chaotic}.
  \newblock {\em Phys. Lett. B}, 788:486--493, 2019.
  \newblock \href {http://arxiv.org/abs/1803.06527} {\path{arXiv:1803.06527}},
  \href {https://doi.org/10.1016/j.physletb.2018.11.050}
  {\path{doi:10.1016/j.physletb.2018.11.050}}.

  \bibitem{Maldacena:2015waa}
  Juan Maldacena, Stephen~H. Shenker, and Douglas Stanford.
  \newblock {A bound on chaos}.
  \newblock {\em JHEP}, 08:106, 2016.
  \newblock \href {http://arxiv.org/abs/1503.01409} {\path{arXiv:1503.01409}},
  \href {https://doi.org/10.1007/JHEP08(2016)106}
  {\path{doi:10.1007/JHEP08(2016)106}}.

  \bibitem{Zhao:2018wkl}
  Qing-Qing Zhao, Yue-Zhou Li, and H.~Lu.
  \newblock {Static Equilibria of Charged Particles Around Charged Black Holes:
    Chaos Bound and Its Violations}.
  \newblock {\em Phys. Rev. D}, 98(12):124001, 2018.
  \newblock \href {http://arxiv.org/abs/1809.04616} {\path{arXiv:1809.04616}},
  \href {https://doi.org/10.1103/PhysRevD.98.124001}
  {\path{doi:10.1103/PhysRevD.98.124001}}.

  \bibitem{Guo:2020pgq}
  Xiaobo Guo, Kangkai Liang, Benrong Mu, Peng Wang, and Mingtao Yang.
  \newblock {Minimal Length Effects on Motion of a Particle in Rindler Space}.
  \newblock {\em Chin. Phys. C}, 45(2):023115, 2021.
  \newblock \href {http://arxiv.org/abs/2007.07744} {\path{arXiv:2007.07744}},
  \href {https://doi.org/10.1088/1674-1137/abcf20}
  {\path{doi:10.1088/1674-1137/abcf20}}.

  \bibitem{PandoZayas:2010xpn}
  Leopoldo~A. Pando~Zayas and Cesar~A. Terrero-Escalante.
  \newblock {Chaos in the Gauge / Gravity Correspondence}.
  \newblock {\em JHEP}, 09:094, 2010.
  \newblock \href {http://arxiv.org/abs/1007.0277} {\path{arXiv:1007.0277}},
  \href {https://doi.org/10.1007/JHEP09(2010)094}
  {\path{doi:10.1007/JHEP09(2010)094}}.

  \bibitem{Ma:2014aha}
  Da-Zhu Ma, Jian-Pin Wu, and Jifang Zhang.
  \newblock {Chaos from the ring string in a Gauss-Bonnet black hole in AdS5
    space}.
  \newblock {\em Phys. Rev. D}, 89(8):086011, 2014.
  \newblock \href {http://arxiv.org/abs/1405.3563} {\path{arXiv:1405.3563}},
  \href {https://doi.org/10.1103/PhysRevD.89.086011}
  {\path{doi:10.1103/PhysRevD.89.086011}}.

  \bibitem{Basu:2016zkr}
  Pallab Basu, Pankaj Chaturvedi, and Prasant Samantray.
  \newblock {Chaotic dynamics of strings in charged black hole backgrounds}.
  \newblock {\em Phys. Rev. D}, 95(6):066014, 2017.
  \newblock \href {http://arxiv.org/abs/1607.04466} {\path{arXiv:1607.04466}},
  \href {https://doi.org/10.1103/PhysRevD.95.066014}
  {\path{doi:10.1103/PhysRevD.95.066014}}.

  \bibitem{Hashimoto:2018fkb}
  Koji Hashimoto, Keiju Murata, and Norihiro Tanahashi.
  \newblock {Chaos of Wilson Loop from String Motion near Black Hole Horizon}.
  \newblock {\em Phys. Rev. D}, 98(8):086007, 2018.
  \newblock \href {http://arxiv.org/abs/1803.06756} {\path{arXiv:1803.06756}},
  \href {https://doi.org/10.1103/PhysRevD.98.086007}
  {\path{doi:10.1103/PhysRevD.98.086007}}.

  \bibitem{Cubrovic:2019qee}
  Mihailo \v{C}ubrovi\'c.
  \newblock {The bound on chaos for closed strings in Anti-de Sitter black hole
    backgrounds}.
  \newblock {\em JHEP}, 12:150, 2019.
  \newblock \href {http://arxiv.org/abs/1904.06295} {\path{arXiv:1904.06295}},
  \href {https://doi.org/10.1007/JHEP12(2019)150}
  {\path{doi:10.1007/JHEP12(2019)150}}.

  \bibitem{Ma:2019ewq}
  Da-Zhu Ma, Dan Zhang, Guoyang Fu, and Jian-Pin Wu.
  \newblock {Chaotic dynamics of string around charged black brane with
    hyperscaling violation}.
  \newblock {\em JHEP}, 01:103, 2020.
  \newblock \href {http://arxiv.org/abs/1911.09913} {\path{arXiv:1911.09913}},
  \href {https://doi.org/10.1007/JHEP01(2020)103}
  {\path{doi:10.1007/JHEP01(2020)103}}.

  \bibitem{Ma:2022tvs}
  Da-Zhu Ma, Fang Xia, Dan Zhang, Guo-Yang Fu, and Jian-Pin Wu.
  \newblock {Chaotic dynamics of string around the conformal black hole}.
  \newblock {\em Eur. Phys. J. C}, 82(4):372, 2022.
  \newblock \href {http://arxiv.org/abs/2205.00226} {\path{arXiv:2205.00226}},
  \href {https://doi.org/10.1140/epjc/s10052-022-10338-5}
  {\path{doi:10.1140/epjc/s10052-022-10338-5}}.

  \bibitem{Cardoso:2008bp}
  Vitor Cardoso, Alex~S. Miranda, Emanuele Berti, Helvi Witek, and Vilson~T.
  Zanchin.
  \newblock {Geodesic stability, Lyapunov exponents and quasinormal modes}.
  \newblock {\em Phys. Rev. D}, 79:064016, 2009.
  \newblock \href {http://arxiv.org/abs/0812.1806} {\path{arXiv:0812.1806}},
  \href {https://doi.org/10.1103/PhysRevD.79.064016}
  {\path{doi:10.1103/PhysRevD.79.064016}}.

  \bibitem{Guo:2021enm}
  Guangzhou Guo, Peng Wang, Houwen Wu, and Haitang Yang.
  \newblock {Quasinormal Modes of Black Holes with Multiple Photon Spheres}.
  \newblock 12 2021.
  \newblock \href {http://arxiv.org/abs/2112.14133} {\path{arXiv:2112.14133}}.

  \bibitem{Frolov:1999pj}
  Andrei~V. Frolov and Arne~L. Larsen.
  \newblock {Chaotic scattering and capture of strings by black hole}.
  \newblock {\em Class. Quant. Grav.}, 16:3717--3724, 1999.
  \newblock \href {http://arxiv.org/abs/gr-qc/9908039}
  {\path{arXiv:gr-qc/9908039}}, \href
  {https://doi.org/10.1088/0264-9381/16/11/316}
  {\path{doi:10.1088/0264-9381/16/11/316}}.

  \bibitem{verner1996high}
  JH~Verner.
  \newblock High-order explicit runge-kutta pairs with low stage order.
  \newblock {\em Applied numerical mathematics}, 22(1-3):345--357, 1996.

\end{thebibliography}
\end{document}